\documentclass[aps,twocolumn,prb,floatfix,superscriptaddress,amsmath,amssymb]{revtex4-2} 
\usepackage{dcolumn}
\usepackage{amsmath}
\usepackage{mathrsfs}
\usepackage{txfonts}
\usepackage{bm}
\usepackage[T1]{fontenc}
\usepackage{xspace}
\usepackage{braket}
\usepackage{bbold}
\usepackage{mathtools}
\newcommand{\mean}[1]{\left<#1\right>}

\newcommand{\bigbraket}[1]{\big\langle #1 \big\rangle}

\newcommand{\red}[1]{\textcolor{black}{#1}}

\usepackage{graphicx}
\usepackage{placeins}
\usepackage{needspace}
\usepackage{hyperref}
\usepackage{color}
\usepackage{xcolor}
\usepackage[version=3]{mhchem}
\hypersetup{
        colorlinks=true,
        citecolor=blue,
        urlcolor=blue,
        linkcolor=blue
}

\renewcommand*{\HyperDestNameFilter}[1]{\jobname-#1}

\begin{document}
\newcommand{\bk}{\bm{k}}
\newcommand{\bS}{\bm{S}}
\newcommand{\kt}{\kappa_{xy}}
\newcommand{\JtwoZero}{0.5215}
\newcommand{\JtwoPi}{0.2945}
\newcommand{\JtwoPiHalf}{0.4000}
\newcommand{\GapSpread}{0.605\%}
\newcommand{\KappaRatioPiZero}{34.3}
\newcommand{\KappaRatioPiHalfZero}{11.6}
\newcommand{\KappaRatioPiHalfPi}{0.339}

\let\emph\textit

\title{Thermal Hall Signatures of Distinct Schwinger-Boson Flux Sectors on the Honeycomb Lattice}

\author{Daiki Sasamoto}
\email[sasamoto.daiki.r6@dc.tohoku.ac.jp]{}
\affiliation{
  Department of Physics, Graduate School of Science, Tohoku University,
  Sendai, Miyagi 980-8578, Japan
}

\date{October 25, 2026}

\begin{abstract}
Thermal transport offers a bulk probe of charge-neutral excitations in
frustrated magnets.  It remains challenging, however, to determine whether
such transport can distinguish different parton mean-field structures of the
same spin system.  This question is particularly relevant on the honeycomb
lattice, where distinct \(0\)-, \(\pi\)-, and chiral \(\pi/2\)-flux sectors
have been proposed.  In this paper, we compare self-consistent saddle-point
solutions in these three sectors for a frustrated \(J_1-J_2\) Heisenberg
model with a next-nearest-neighbor Dzyaloshinskii--Moriya interaction and a
perpendicular magnetic field.  Using
finite-temperature Schwinger-boson mean-field theory and the Kubo formula for
bosonic Bogoliubov--de Gennes systems, we evaluate the intrinsic spinon
thermal Hall conductivity.  At a
common parameter point where all three branches remain gapped, the \(0\)-flux
response is positive and the \(\pi\)-flux response is negative for \(D>0\)
and \(h>0\) under the sign conventions used in this paper, whereas the
\(\pi/2\)-flux response changes sign with increasing temperature.  We further
show that a projective
\(\widetilde C_6\mathcal T\) symmetry forces the zero-field response of a
fixed chiral \(\pi/2\)-flux domain to vanish.  Within the mean-field regime
examined in this paper, the sign and temperature dependence of \(\kappa_{xy}/T\)
therefore provide a flux-sensitive transport signature.
\end{abstract}

\maketitle

\section{Introduction}
\label{sec:introduction}

Strongly correlated electron systems host collective states that cannot be
understood by treating interactions as a weak correction to an
independent-electron band structure.  When the Coulomb repulsion competes with the
electronic bandwidth, charge motion can be suppressed while local spin degrees
of freedom remain active, providing the microscopic setting for Mott insulating
magnets \cite{Hubbard-1963,Imada-1998}.  More radically, interactions can
reorganize a many-electron system into emergent excitations carrying only part
of the quantum numbers of its microscopic constituents, as demonstrated by the
fractionally charged excitations of the fractional quantum Hall effect
\cite{Tsui-Stormer-Gossard-1982,Laughlin-1983}.  The
resonating-valence-bond proposal extended this principle to Mott magnets
\cite{Anderson-1973}.  A quantum spin liquid can avoid conventional magnetic
order and instead exhibit long-range entanglement, emergent gauge structure,
and fractionalized quasiparticles
\cite{Balents-2010,Savary-Balents-2016,Zhou-Kanoda-2017,
Broholm-Cava-2020}.  In a parton
description, electrically neutral spinons carry spin and propagate through an
emergent gauge background \cite{Wen-2002,Kitaev-2006}.  When time-reversal
symmetry is broken, their bands can acquire nonzero Berry curvature and integer
Chern numbers, connecting chiral spin liquids to fractional quantum Hall
physics \cite{Kalmeyer-Laughlin-1987,Wen-1989,Kitaev-2006}.  The absence of a
local order parameter makes such states difficult to identify experimentally
\cite{Knolle-Moessner-2019,Wen-Yu-2019}, and the charge neutrality of spinons
makes heat transport one of the few bulk transport probes to which they can
contribute directly \cite{Yamashita-2020,Hirschberger-Krizan-2015}.  Thermal
Hall signals attributed to magnetic excitations have also been reported in
kagome magnets
\cite{Hirschberger-Chisnell-2015,Watanabe-2016,Doki-Akazawa-2018}.
Magnon thermal Hall transport in ordered magnets provides a well-established
bosonic-band example of the same Berry-curvature mechanism considered here.
The thermally weighted response can even change sign with temperature
\cite{Katsura-Nagaosa-Lee-2010,Onose-2010,Matsumoto-Murakami-2011,
Matsumoto-Murakami-2011-2,Shindou-Matsumoto-Murakami-Ohe-2013,
Mook-Henk-Mertig-2014,Fujiwara-Kitamura-Morimoto-2022,McClarty-2022,
Zhang-Gao-Chen-2024}.  Thermal Hall transport has also been used to probe
neutral topological excitations, most prominently in the Kitaev-material
candidate \(\alpha\)-\(\mathrm{RuCl}_3\)
\cite{Kasahara-2018,Yokoi-2021,Bruin-2022}.  The interpretation of such
signals requires care because phonons can also carry a transverse heat current
\cite{Akazawa-2020,Lefrancois-2022}, and because an integer Chern number of a
generic bosonic band does not imply a quantized plateau.  The conductivity
\(\kappa_{xy}\) is
determined by a temperature-dependent Bose weighting of the Berry curvature,
which itself contains interband matrix elements
\cite{Matsumoto-Murakami-2011-2}.

Against this background, Schwinger-boson mean-field theory (SBMFT) provides a
natural parton description of bosonic spinons
\cite{Arovas-Auerbach-1998,Read-Sachdev-1991,Sachdev-Read-1991,Sachdev-1992,
Sarker-Jayaprakash-Krishnamurthy-Ma-1989,Trumper-Manuel-1997,
Flint-Coleman-2009,Merino-Holt-Powell-2014}.  Its
projective-symmetry-group (PSG) structure distinguishes Ansätze that realize
physical lattice symmetries only after an accompanying gauge transformation
\cite{Wen-2002,Wang-Vishwanath-2006}.  Chiral PSG constructions allow
time-reversal-breaking fluxes
\cite{Messio-Lhuillier-2013,Bieri-Lhuillier-Messio-2016}.  On the honeycomb
lattice, a symmetry-preserving PSG classification identified
time-reversal-invariant \(0\)- and \(\pi\)-flux Schwinger-boson states
\cite{Wang-2010}, while an eight-site chiral singlet \(\pi/2\)-flux Ansatz was
subsequently introduced for the integer-spin Heisenberg--Kitaev model
\cite{Ralko-Merino-2024}.  This chiral state is distinct from the
time-reversal-invariant PSG classes of Ref.~\cite{Wang-2010}.  In that
Kitaev-magnet setting, the \(\pi/2\)-flux Ansatz reproduces qualitative
features of the exact-diagonalization dynamical structure factor, including
nearly dispersionless low-energy features, and
remains gapped within SBMFT up to \(S\lesssim2\)
\cite{Ralko-Merino-2024,Sasamoto-Ralko-Merino-Nasu-2026}.  Here we extend the corresponding honeycomb
\(\pi/2\)-flux construction to a frustrated antiferromagnetic
\(J_1-J_2\) Heisenberg model and compare it directly with the \(0\)- and
\(\pi\)-flux states at the same microscopic couplings.

Mean-field free energies can be used to compare stationary solutions within a
fixed decoupling scheme.  In SBMFT, however, the local boson-number constraint
of the physical spin Hilbert space is imposed only on average, so these free
energies are not rigorous variational upper bounds for the original spin
Hamiltonian.  We therefore do not use their ordering to construct a phase
diagram.  Instead, we ask how intrinsic thermal transport would
distinguish each self-consistent gapped branch.
Previous parton calculations have studied spinon thermal Hall
responses
\cite{Lee-Han-Lee-2015,Park-Yang-2020,Samajdar-Chatterjee-2019,
Teng-Zhang-Samajdar-Scheurer-Sachdev-2020}.  On the honeycomb lattice, an
SBMFT study investigated thermal Hall transport in a
\(J_1-J_2-J_\chi\) model and characterized a chiral \(\mathbb Z_2\) phase
\cite{Mukherjee-2023}.  Here, by contrast, we compare self-consistent \(0\)-,
\(\pi\)-, and chiral \(\pi/2\)-flux branches of a \(J_1-J_2\) Heisenberg model
with a next-nearest-neighbor Dzyaloshinskii--Moriya interaction and a
perpendicular magnetic field, using identical microscopic couplings for all
three branches.
The gap controls the thermal activation scale, but not the full spinon
dispersion, paraunitary Bogoliubov eigenvectors, Berry-curvature distribution,
or interband matrix elements entering the Kubo response.  These additional
ingredients can change both the sign and scale of \(\kappa_{xy}\).

We obtain self-consistent solutions for these three flux sectors and follow
each branch as temperature and couplings are varied.  We then evaluate the
intrinsic thermal Hall response using the bosonic Bogoliubov--de Gennes (BdG)
Kubo formula.  We find
that, at identical microscopic couplings, the
three sectors exhibit qualitatively different odd-in-field thermal Hall
responses.  At the common parameter point, the \(0\)- and \(\pi\)-flux
responses have opposite signs, while the \(\pi/2\)-flux response changes sign
as the temperature increases.  We therefore propose the combined sign and temperature
dependence of \(\kappa_{xy}\), rather than its magnitude at a single
temperature, as a flux-sensitive diagnostic within the parameter regime
studied.

This paper is organized as follows.  In Sec.~\ref{sec:model}, we introduce the
honeycomb \(J_1-J_2\) model and define the Dzyaloshinskii--Moriya and
magnetic-field conventions.  Section~\ref{sec:method} presents the
Schwinger-boson mean-field formulation, the three flux Ansätze, and the
formalism for calculating the intrinsic thermal Hall conductivity.  In
Sec.~\ref{sec:results}, we present the common-parameter comparison, the
coupling and field dependences, and the band-resolved Berry-curvature
distributions.  Section~\ref{sec:discussion} discusses the resulting
flux-sensitive transport signature, the zero-field cancellation in the
chiral flux state, and the complementary information provided by the
dynamical structure factor.  Section~\ref{sec:summary} summarizes our findings.
The Supplemental Material provides the complete Ansatz specification, the
energy-resolved analysis of the \(\pi/2\)-flux sign reversal,
the parameter-range and convergence checks, an ordered-state thermal Hall benchmark, and the
formulation and results for the dynamical structure
factor~\cite{SupplementalMaterial}.

\section{Honeycomb \texorpdfstring{$J_1-J_2$}{J1-J2} Model}
\label{sec:model}

\begin{figure}[t]
\centering
\includegraphics[width=\columnwidth]{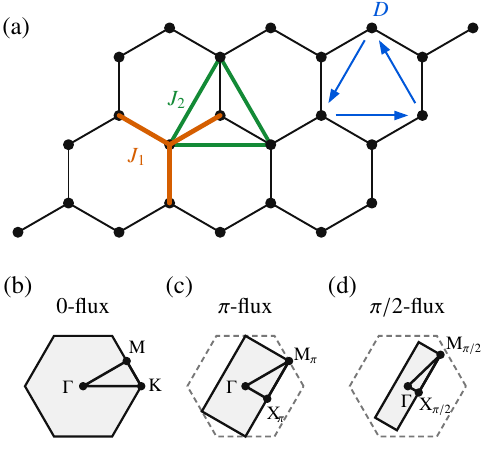}
\caption{
(a) Honeycomb lattice with three representative nearest-neighbor \(J_1\)
bonds shown in orange.  Green and blue mark the NNN \(J_2\) coupling and DM
orientation, respectively.  They are displayed on distinct representative
triangles for clarity, although both act on every NNN bond.  A blue arrow
from \(i\) to \(j\) defines \(\nu_{ij}=+1\).
(b) Physical first Brillouin zone and the
\(\Gamma\)-\(\mathrm M\)-\(\mathrm K\)-\(\Gamma\) path used for the
\(0\)-flux sector.
(c),(d) Magnetic reciprocal primitive cells and the
\(\Gamma\)-\(\mathrm X_{\pi}\)-\(\mathrm M_{\pi}\)-\(\Gamma\) and
\(\Gamma\)-\(\mathrm X_{\pi/2}\)-\(\mathrm M_{\pi/2}\)-\(\Gamma\) paths,
respectively.  The subscripts distinguish high-symmetry points of the two
magnetic zones, and the dashed hexagon marks the physical first Brillouin zone.
}
\label{fig:model_bz}
\end{figure}

We consider the spin-\(1/2\) \(J_1-J_2\) Heisenberg model on the honeycomb
lattice shown in Fig.~\ref{fig:model_bz}(a), described by the Hamiltonian
\begin{align}
\mathcal{H}={}&
J_1\sum_{\langle ij\rangle}\bS_i\cdot\bS_j
+J_2\sum_{\langle\!\langle ij\rangle\!\rangle}\bS_i\cdot\bS_j
\notag\\
&+D\sum_{\langle\!\langle ij\rangle\!\rangle}
\nu_{ij}\hat{\bm z}\cdot(\bS_i\times\bS_j)
-h\sum_i S_i^z ,
\label{eq:hamiltonian}
\end{align}
where \(\langle ij\rangle\) and \(\langle\!\langle ij\rangle\!\rangle\)
denote nearest-neighbor and next-nearest-neighbor (NNN) bonds, respectively,
and \(h\) is the Zeeman energy for a field along \(+\hat{\bm z}\).  We take
antiferromagnetic \(J_1,J_2>0\).  The coupling \(J_1\) connects nearest-neighbor
sites on opposite sublattices and favors bipartite N\'eel correlations, whereas
\(J_2\) connects NNN sites on the same sublattice and introduces magnetic
frustration.
Numerical studies of the
spin-\(1/2\) \(J_1-J_2\) model find an intermediate nonmagnetic regime,
but disagree on whether it is best described as a spin liquid, weak
plaquette-valence-bond order, or a narrow regime separating N\'eel and
valence-bond phases
\cite{Mulder-Ganesh-Capriotti-Paramekanti-2010,Cabra-Lamas-2011,
Clark-Abanin-Sondhi-2011,Albuquerque-2011,Mosadeq-2011,
Mezzacapo-Boninsegni-2012,Zhu-Huse-White-2013,
Gong-Sheng-Motrunich-Fisher-2013,Ferrari-Bieri-Becca-2017}.  The same model has
been studied within SBMFT since early calculations of its excitation spectrum
and magnetic response \cite{Mattsson-Frojdh-1994}.  Subsequent self-consistent
Schwinger-boson studies consistently found an intermediate magnetically
disordered regime with a finite bosonic excitation gap, while differing in its
upper boundary and in the ordered phase found at larger \(J_2/J_1\)
\cite{Zhang-Lamas-2013,Yu-Liu-2014,Merino-Ralko-2018}.  These differences among
mean-field results motivate a systematic comparison of symmetry-compatible
Ans\"atze spanning distinct flux sectors.

The third term in Eq.~\eqref{eq:hamiltonian} is an NNN
Dzyaloshinskii--Moriya (DM) interaction
\cite{Dzyaloshinskii-1958,Moriya-1960}.  On every NNN bond we define
\begin{equation}
\bm D_{ij}=D\nu_{ij}\hat{\bm z},
\qquad \nu_{ji}=-\nu_{ij},\quad \nu_{ij}=\pm1 .
\label{eq:dm_vector}
\end{equation}
Thus the DM interaction acts on NNN bonds within each sublattice.  Its
orientation on the honeycomb lattice is shown in
Fig.~\ref{fig:model_bz}(a).  A
bond directed from the tail to the head of a blue arrow in
Fig.~\ref{fig:model_bz}(a) has \(\nu_{ij}=+1\).  The oppositely directed bond has
\(\nu_{ij}=-1\), and the pattern is repeated by lattice translations.  This
out-of-plane DM interaction reduces spin-rotation symmetry from
SU($2$) to U($1$) about \(\hat{\bm z}\), but is itself even under time reversal.
The Zeeman term breaks time reversal explicitly.  Reversing every DM arrow is
equivalent to \(D\to-D\), which reverses the sign of the thermal Hall response
while leaving the spectrum unchanged.
Throughout this paper, we set \(\hbar=k_{\mathrm B}=1\), use \(J_1\) as the
energy unit with \(J_1=1\), and set the nearest-neighbor distance to \(a=1\).
The real- and reciprocal-space conventions, including the physical and
magnetic Brillouin zones and the high-symmetry paths used below, are
summarized in Fig.~\ref{fig:model_bz}.  Exact vector and point coordinates are
listed in the Supplemental Material~\cite{SupplementalMaterial}.

\section{Method}
\label{sec:method}

\subsection{Schwinger-Boson Mean-Field Theory}
\label{sec:sbmft}

Schwinger-boson mean-field theory rewrites a spin as a bilinear of bosonic
spinons.  At each site \(i\), we introduce two canonical bosons
\(b_{i\mu}\), with \(\mu=\uparrow,\downarrow\), satisfying
\([b_{i\mu},b_{j\nu}^{\dagger}]=\delta_{ij}\delta_{\mu\nu}\), and represent
the spin operator as
\begin{equation}
 S_i^{\gamma}=\frac12
 \sum_{\mu,\nu}b_{i\mu}^{\dagger}\sigma^{\gamma}_{\mu\nu}b_{i\nu},
 \qquad \gamma=x,y,z.
 \label{eq:schwinger_boson}
\end{equation}
Here \(\sigma^{\gamma}\) are the Pauli matrices, and \(\gamma\) labels the
physical spin component.
Equation~\eqref{eq:schwinger_boson}
is an exact operator representation only after projecting the enlarged
bosonic Fock space back to the physical spin Hilbert space.  The auxiliary
bosons allow arbitrary local occupation numbers, whereas a spin-\(S\)
degree of freedom is obtained by imposing the local constraint
\begin{equation}
 n_i\equiv\sum_{\mu}b_{i\mu}^{\dagger}b_{i\mu}=2S .
 \label{eq:local_constraint}
\end{equation}
The states satisfying \(n_i=2S\) span the physical spin-\(S\) multiplet at
site \(i\).  At the mean-field level, this exact operator condition is imposed
on expectation values.  For every inequivalent site in the magnetic unit
cell, we require
\begin{equation}
 \mean{n_i}=\kappa.
 \label{eq:mean_number_constraint}
\end{equation}
These constraints are enforced by Lagrange multipliers in the quadratic
Hamiltonian
\cite{Arovas-Auerbach-1998,Read-Sachdev-1991,Sachdev-Read-1991,Sachdev-1992}.
Symmetry-equivalent sites share the same constraint and multiplier.  The
independent Lagrange multipliers for each Ansatz are specified in the
Supplemental Material~\cite{SupplementalMaterial}.
For a spin-\(1/2\) system, the microscopic boson-number constraint corresponds
to \(\kappa=2S=1\).  We instead use \(\kappa=\sqrt{3}-1\), a standard
spin-length correction that restores the spin-\(1/2\) equal-time local-moment
sum rule within the Gaussian mean-field treatment
\cite{Mezio-2011,Messio-Bernu-2012,Lugan-Jaubert-2022}.
\label{para:constraint_prescription_start}
\red{This is an effective local-moment normalization prescription for
uncondensed mean-field states.  It does not enforce the exact local
boson-number constraint, which remains imposed only on average.}
\label{para:constraint_prescription_end}

The representation in Eq.~\eqref{eq:schwinger_boson} has a local U($1$) gauge
redundancy.  The transformation
\(b_{i\mu}\to e^{i\theta_i}b_{i\mu}\) leaves every physical spin operator
unchanged but transforms the bond mean fields.  Individual bond amplitudes are
therefore gauge dependent, whereas the loop variable defined in
Sec.~\ref{sec:mean_field_ansatzes} is gauge invariant.

The interaction is first expressed in terms of bond channels.  Following
generalized Schwinger-boson formulations of anisotropic spin interactions
\cite{Ghioldi-Mezio-Manuel-Singh-Oitmaa-Trumper-2015,Ralko-Merino-2024,
Taran-Ralko-Temnikov-Mazurenko-Streltsov-Iqbal-2026,
Sasamoto-Nasu-2025,Sasamoto-Nasu-2026}, the SU($2$)-invariant
singlet-pairing and hopping channels are supplemented by SU($2$)-breaking
spin-vector bond operators.  For the present Heisenberg and out-of-plane DM
Hamiltonian it is sufficient to retain only their \(z\) components.  On a directed bond
\((i,j)\), we define
\begin{subequations}
\label{eq:bond_operators}
\begin{align}
 \mathcal A_{ij}
 &=\frac12\left(
 b_{i\uparrow}b_{j\downarrow}
 -b_{i\downarrow}b_{j\uparrow}
 \right),
 \\
 \mathcal B_{ij}
 &=\frac12\left(
 b_{i\uparrow}^{\dagger}b_{j\uparrow}
 +b_{i\downarrow}^{\dagger}b_{j\downarrow}
 \right),
 \\
 \mathcal C^z_{ij}
 &=\frac12\left(
 b_{i\uparrow}^{\dagger}b_{j\uparrow}
 -b_{i\downarrow}^{\dagger}b_{j\downarrow}
 \right),
 \\
 \mathcal D^z_{ij}
 &=-\frac12\left(
 b_{i\uparrow}b_{j\downarrow}
 +b_{i\downarrow}b_{j\uparrow}
 \right).
\end{align}
\end{subequations}
Here \(\mathcal A_{ij}\) and \(\mathcal D^z_{ij}\) are the spin-singlet and
\(S^z=0\) triplet pairing channels, respectively, whereas
\(\mathcal B_{ij}\) and \(\mathcal C^z_{ij}\) are the spin-scalar and
\(z\)-component spin-vector hopping channels.  The spin-vector channels
\(\mathcal C^z_{ij}\) and \(\mathcal D^z_{ij}\) are absent in a purely
SU($2$)-symmetric Heisenberg decoupling but become symmetry-allowed, and
generally necessary, once the out-of-plane DM term lowers the spin-rotation
symmetry from SU($2$) to U($1$).  Together, the four bond operators give the exact identities
\begin{align}
 \bS_i\!\cdot\!\bS_j
 &= \mathopen{:}\,\mathcal B_{ij}^{\dagger}\mathcal B_{ij}\,\mathclose{:}
   -\mathcal A_{ij}^{\dagger}\mathcal A_{ij},
 \label{eq:exchange_bonds}\\
 \hat{\bm z}\!\cdot\!(\bS_i\!\times\!\bS_j)
 &=\frac{i}{2}\bigl(
 \mathcal D_{ij}^{z\dagger}\mathcal A_{ij}
 -\mathcal A_{ij}^{\dagger}\mathcal D^z_{ij}
 -\mathopen{:}\,\mathcal C_{ij}^{z\dagger}\mathcal B_{ij}\,\mathclose{:}
 +\mathopen{:}\,\mathcal B_{ij}^{\dagger}\mathcal C^z_{ij}\,\mathclose{:}\bigr).
 \label{eq:dm_bonds}
\end{align}
The notation \(\mathopen{:}\mathcal O\mathclose{:}\) denotes bosonic normal
ordering.  All creation operators in \(\mathcal O\) are placed to the left of
all annihilation operators.  We decouple each
product of bond operators appearing explicitly in
Eqs.~\eqref{eq:exchange_bonds} and \eqref{eq:dm_bonds} according to
\begin{equation}
 \mathcal X_{ij}^{\dagger}\mathcal Y_{ij}\simeq
 \mean{\mathcal X_{ij}^{\dagger}}\mathcal Y_{ij}
 +\mathcal X_{ij}^{\dagger}\mean{\mathcal Y_{ij}}
 -\mean{\mathcal X_{ij}^{\dagger}}\mean{\mathcal Y_{ij}},
 \label{eq:mf_decoupling}
\end{equation}
where \(\mathcal X_{ij}\) and \(\mathcal Y_{ij}\) denote any of the four bond
operators defined in Eq.~\eqref{eq:bond_operators}.  For the hopping-type terms
in Eqs.~\eqref{eq:exchange_bonds} and \eqref{eq:dm_bonds}, the decoupling in
Eq.~\eqref{eq:mf_decoupling} is applied to the normal-ordered products shown
there.  Their complex expectation values are the bond mean fields.  The constraint term is
correspondingly
\begin{equation}
 \mathcal H_{\lambda}
 =\sum_i\lambda_i(n_i-\kappa),
 \label{eq:constraint_term}
\end{equation}
and is added to this decoupled Hamiltonian together with the Zeeman term.
The multipliers \(\lambda_i\) are identical on symmetry-equivalent sites.  The
independent set used for each Ansatz is given in the Supplemental Material~\cite{SupplementalMaterial}.
We label each site as \(i=(l,m)\), where \(N\) is the total number of lattice
sites, \(l=1,\ldots,N/M\) labels the real-space magnetic unit cells, and
\(m=1,\ldots,M\) labels the sites inside the magnetic cell.  The value of \(M\)
is fixed by the Ansatz.  In real space we introduce the cell Nambu spinor
\begin{align}
 \Psi_l^{\dagger}
 =\Bigl(
 &b_{l,1,\uparrow}^{\dagger},\ldots,
 b_{l,M,\uparrow}^{\dagger},
 b_{l,1,\downarrow}^{\dagger},\ldots,
 b_{l,M,\downarrow}^{\dagger},
 \notag\\
 &b_{l,1,\uparrow},\ldots,
 b_{l,M,\uparrow},
 b_{l,1,\downarrow},\ldots,
 b_{l,M,\downarrow}
 \Bigr).
 \label{eq:real_space_nambu_spinor}
\end{align}
The decoupled mean-field Hamiltonian can then be written as a real-space
bosonic BdG Hamiltonian,
\begin{equation}
 \mathcal H_{\mathrm{MF}}
 =\frac12\sum_{l,l'}\Psi_l^{\dagger}
 \mathcal M_{ll'}\Psi_{l'}+\mathrm{const.}
 \label{eq:real_space_bdg_form}
\end{equation}
The additive constant must be retained when constructing the saddle-point
free energy.  Once a self-consistent saddle-point solution has been fixed,
however, it does
not enter the BdG matrix and therefore affects neither the thermal Hall
response nor the dynamical spin correlations.  We omit it henceforth.
Here \(\mathcal M_{ll'}\) is a \(4M\times4M\) block in Nambu space.
Its normal blocks contain the hopping-type mean fields, Lagrange multipliers,
and Zeeman terms, while its anomalous blocks contain the pairing-type mean
fields.

We Fourier transform the bosons as
\begin{align}
 b_{l m\mu}
 &=\sqrt{\frac{M}{N}}
   \sum_{\bk} e^{i\bk\cdot\bm R_l} b_{\bk m\mu},
 \notag\\
 b_{\bk m\mu}
 &=\sqrt{\frac{M}{N}}
   \sum_l e^{-i\bk\cdot\bm R_l} b_{l m\mu},
 \label{eq:fourier_transform}
\end{align}
where \(\bm R_l\) is the magnetic-cell position and \(\bk\) is summed over
the magnetic Brillouin zone (MBZ).  The momentum-space Nambu spinor is
\begin{align}
 \Psi_{\bk}^{\dagger}
 =\Bigl(
 &b_{\bk,1,\uparrow}^{\dagger},\ldots,
 b_{\bk,M,\uparrow}^{\dagger},
 b_{\bk,1,\downarrow}^{\dagger},\ldots,
 b_{\bk,M,\downarrow}^{\dagger},
 \notag\\
 &b_{-\bk,1,\uparrow},\ldots,
 b_{-\bk,M,\uparrow},
 b_{-\bk,1,\downarrow},\ldots,
 b_{-\bk,M,\downarrow}
 \Bigr).
 \label{eq:nambu_spinor}
\end{align}
With the convention of Eq.~\eqref{eq:fourier_transform}, the Hamiltonian takes
the bosonic BdG form
\begin{equation}
 \mathcal H_{\mathrm{MF}}=\frac12\sum_{\bk}\Psi_{\bk}^{\dagger}
 \mathcal M_{\bk}\Psi_{\bk}.
 \label{eq:mf_bdg_form}
\end{equation}
The Fourier-space BdG block is
\begin{equation}
 \mathcal M_{\bk}
 =\sum_{l'}\mathcal M_{ll'}
 e^{-i\bk\cdot(\bm R_l-\bm R_{l'})}
\label{eq:mk_definition}
\end{equation}
Spatial translation symmetry restricts \(\mathcal M_{ll'}\) to depend only on
\(\bm R_l-\bm R_{l'}\).  Equation~\eqref{eq:mk_definition} is therefore independent
of the reference-cell label \(l\).
The bosonic commutation relations of the Nambu spinor are encoded in the metric
\begin{equation}
 \sigma_3=
 \begin{pmatrix}
  \bm{1}_{2M} & 0\\
  0 & -\bm{1}_{2M}
 \end{pmatrix},
 \qquad
 [\Psi_{\bk,\alpha},\Psi_{\bk,\beta}^{\dagger}]
 =(\sigma_3)_{\alpha\beta}.
 \label{eq:sigma3_metric}
\end{equation}
Here \(\bm{1}_{2M}\) is the \(2M\times2M\) identity matrix.  Hence
\(\sigma_3\) acts in the full \(4M\)-dimensional Nambu space.
Following the standard bosonic BdG procedure
\cite{Colpa-1978,Kondo-Akagi-Katsura-2020}, we introduce
quasiparticle operators through
\begin{equation}
 \Psi_{\bk}=T_{\bk}\Gamma_{\bk},
 \qquad
 \Gamma_{\bk}^{\dagger}
 =(\gamma_{\bk,1}^{\dagger},\ldots,\gamma_{\bk,2M}^{\dagger},
   \gamma_{-\bk,1},\ldots,\gamma_{-\bk,2M}).
 \label{eq:bogoliubov_transform}
\end{equation}
The transformation must be paraunitary in order to preserve the bosonic
commutation relations.  It therefore satisfies
\begin{equation}
 T_{\bk}^{\dagger}\sigma_3T_{\bk}
 =T_{\bk}\sigma_3T_{\bk}^{\dagger}
 =\sigma_3 .
 \label{eq:paraunitary_condition}
\end{equation}
Equivalently,
\(T_{\bk}^{-1}=\sigma_3T_{\bk}^{\dagger}\sigma_3\).  We label all
\(4M\) columns of \(T_{\bk}\), including both particle and hole subspaces, by
\(\eta=1,\ldots,4M\) and write them as
\(\lvert u_{\eta\bk}\rangle\).  The generalized eigenvalue problem is
\begin{equation}
 \sigma_3\mathcal M_{\bk}T_{\bk}
 =T_{\bk}\sigma_3\mathcal E_{\bk},
 \label{eq:bosonic_bdg_eigenproblem}
\end{equation}
where, with the ordering in Eq.~\eqref{eq:bogoliubov_transform}, the same
paraunitary matrix gives
\begin{align}
 T_{\bk}^{\dagger}\mathcal M_{\bk}T_{\bk}
 &=\mathcal E_{\bk},
 \notag\\
 \mathcal E_{\bk}
 &=\operatorname{diag}
 \left(
 \varepsilon_{\bk,1},\ldots,\varepsilon_{\bk,2M},
 \varepsilon_{-\bk,1},\ldots,\varepsilon_{-\bk,2M}
 \right).
 \label{eq:bdg_diagonalization}
\end{align}
We write \(\sigma_{3,\eta}\equiv(\sigma_3)_{\eta\eta}\).  Since both
\(\sigma_3\) and \(\mathcal E_{\bk}\) are diagonal, the signed
dynamical-matrix eigenvalue associated with column
\(\lvert u_{\eta\bk}\rangle\) is
\((\sigma_3\mathcal E_{\bk})_{\eta\eta}\).
For a stable noncondensed saddle-point solution,
the positive-norm particle modes have \(\sigma_{3,n}=+1\) and
\((\sigma_3\mathcal E_{\bk})_{nn}=\varepsilon_{n\bk}>0\), whereas the
negative-norm hole partners have
\((\sigma_3\mathcal E_{\bk})_{mm}<0\).  The positive-norm,
positive-energy eigenmodes define the bosonic spinon bands.  At a given
temperature, their Bose occupation factors determine the thermal covariances.
The mean fields and Lagrange multipliers are then updated until the bond
expectation values and the number constraints in
Eq.~\eqref{eq:mean_number_constraint} are satisfied self-consistently.
The flux Ansätze and their gauge-invariant classification are specified
in Sec.~\ref{sec:mean_field_ansatzes}.

\subsection{Mean-field Ansätze}
\label{sec:mean_field_ansatzes}

\begin{figure}[t]
\centering
\includegraphics[width=\columnwidth]{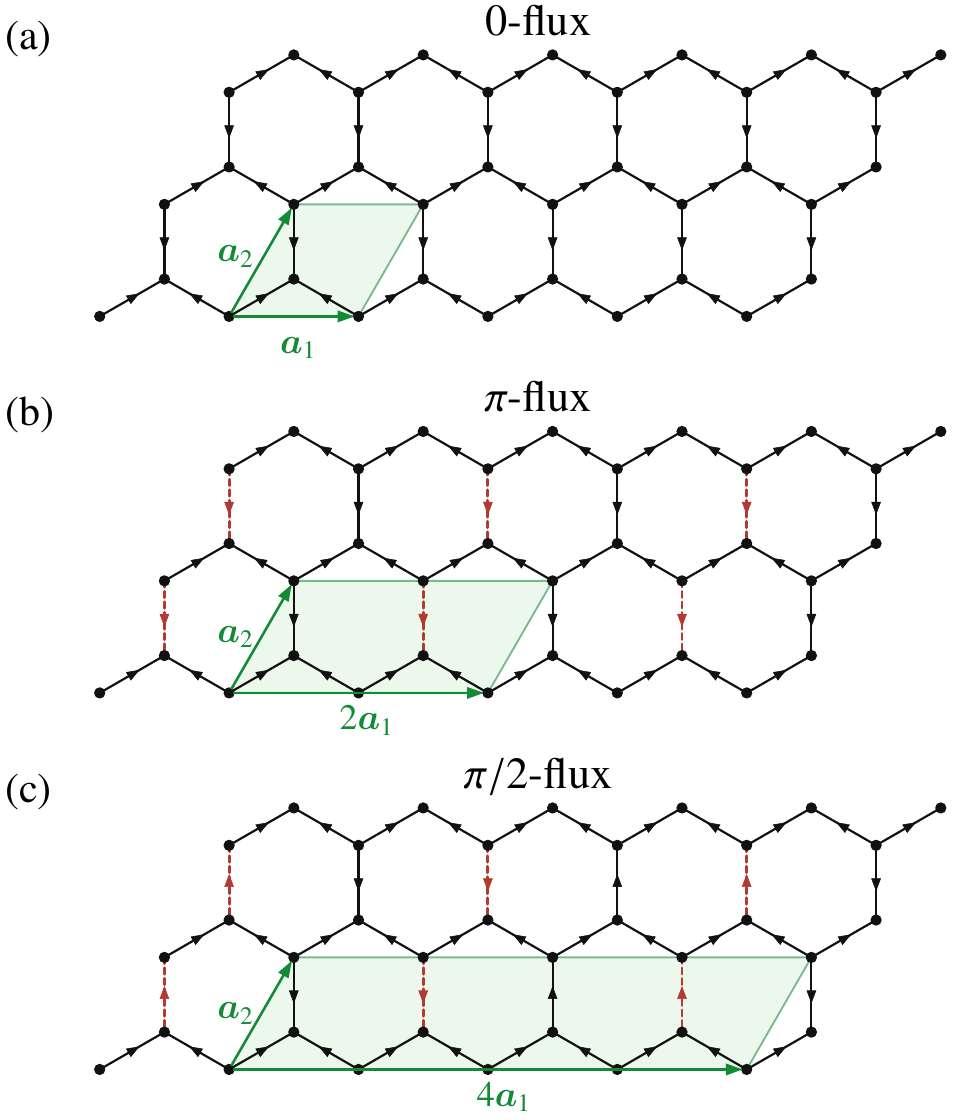}
\caption{
Mean-field Ansätze and gauge conventions on the honeycomb lattice.
(a) \(0\)-flux representative with phase \(1\) on every nearest-neighbor bond.
(b) \(\pi\)-flux representative in a \(2\bm a_1\times\bm a_2\) magnetic
cell.  Red dashed nearest-neighbor bonds have an additional factor \(-1\).
(c) \(\pi/2\)-flux representative in the
\(4\bm a_1\times\bm a_2\) magnetic cell.  Relative to the displayed bond
orientations, every red dashed bond carries the same additional factor \(i\),
whereas every black bond has unit phase.  Reversing an arrow uses
\(\mathcal A_{ji}=-\mathcal A_{ij}\).  The resulting Wilson phase is
\(\phi_p=\pi/2\) on every hexagon.  Arrowheads at bond centers define the
orientation of \(\mathcal A_{ij}\) and the other mean fields.  Green arrows denote the
magnetic translation vectors, and the shaded regions indicate the magnetic
unit cells.
}
\label{fig:ansatz}
\end{figure}

We next specify the three Schwinger-boson flux Ansätze compared in this
work.  In a PSG construction, lattice symmetries may be implemented only up to
gauge transformations, and distinct mean-field representatives can therefore
share the same microscopic Hamiltonian while carrying different gauge-invariant
Wilson-loop phases
\cite{Wen-2002,Wang-Vishwanath-2006,Wang-2010}.  On nearest-neighbor bonds we write
\(\bigbraket{\mathcal A_{ij}}=A_1\eta_{ij}\), with the oriented pair \(i\to j\)
following the A-to-B bond-center arrowheads in Fig.~\ref{fig:ansatz}.  To define the flux
unambiguously, label the six sites of a hexagon counterclockwise as
\(i_1,\ldots,i_6\) and choose \(i_1\) on the A sublattice.  The singlet
Wilson loop is
\begin{equation}
 \begin{aligned}
 W_p\equiv{}&
 \mean{\mathcal A_{i_1i_2}}
 \left(-\mean{\mathcal A_{i_2i_3}}^{*}\right)
 \mean{\mathcal A_{i_3i_4}}\\[-2pt]
 &\times
 \left(-\mean{\mathcal A_{i_4i_5}}^{*}\right)
 \mean{\mathcal A_{i_5i_6}}
 \left(-\mean{\mathcal A_{i_6i_1}}^{*}\right)
 =\lvert W_p\rvert e^{i\phi_p}.
 \end{aligned}
 \label{eq:singlet_wilson_loop}
\end{equation}
The alternating conjugation and minus signs follow the directed singlet-pairing
convention, including
\(\mathcal A_{ji}=-\mathcal A_{ij}\).  Under
\(b_{i\mu}\to e^{i\theta_i}b_{i\mu}\), every site phase in
Eq.~\eqref{eq:singlet_wilson_loop} appears once with each sign and cancels.
Thus \(\phi_p=\arg W_p\), defined modulo \(2\pi\), is gauge invariant.  The
counterclockwise traversal and A-sublattice starting point are part of the sign
convention.  Reversing either convention complex conjugates \(W_p\).
This construction is the pairing-field Wilson loop used to label the Ansatz,
not a circulation of physical bond currents
\cite{Wang-2010,Messio-Lhuillier-2013,Ralko-Merino-2024,
Sasamoto-Nasu-2025}.  The sectors considered below have
\(\phi_p=0\), \(\pi\), and \(\xi\pi/2\), respectively, where
\(\xi=\pm1\) labels the two conjugate \(\pi/2\)-flux domains.  The \(0\)- and
\(\pi\)-flux states correspond to the time-reversal-symmetric honeycomb
Schwinger-boson PSG classes discussed in Ref.~\cite{Wang-2010}.  A chiral
\(\pi/2\)-flux singlet state was later introduced in the context of
integer-spin Heisenberg--Kitaev magnets \cite{Ralko-Merino-2024}.  The
mean-field construction itself is not restricted to a Kitaev-dominant
Hamiltonian.  We apply this flux pattern to the frustrated \(J_1-J_2\)
Heisenberg model defined in Sec.~\ref{sec:model}.

Figure~\ref{fig:ansatz} summarizes the three gauge representatives.  All
nearest-neighbor phases are \(1\) in the \(0\)-flux state.  Red dashed bonds
carry an additional factor \(-1\) in the \(\pi\)-flux state.  In the
\(\pi/2\)-flux representative, the red dashed bonds all carry the common
factor \(i\) with respect to the displayed bond arrows.  Reexpressing every
bond in the common A-to-B orientation gives the numerical gauge
\(\eta_{m,-\bm a_2}=i^m\), \(m=0,\ldots,3\), with unit phase in the other
two nearest-neighbor directions.  The alternating product on every hexagon is
then \(i^{m+1}(i^m)^*=i\).  The three representatives use
\(\bm a_1\times\bm a_2\), \(2\bm a_1\times\bm a_2\), and
\(4\bm a_1\times\bm a_2\) magnetic cells, respectively.  The first two representatives are
closed under \(\widetilde C_6\), whereas a fixed \(\pi/2\)-flux domain is
closed only under \(\widetilde C_3=\widetilde C_6^2\).  The operation \(\widetilde C_6\)
exchanges the two conjugate chiral domains.
Every bond amplitude generated from a chosen independent representative by
magnetic translations and the retained projective point-group operations is
included with its symmetry-fixed phase or complex conjugation.  This does not
mean that all four bond channels are varied independently on every bond.  The
independent saddle-point variables
and the explicit rules that generate their complete sets of symmetry-related bonds are given in
the Supplemental Material~\cite{SupplementalMaterial}.

\subsection{Thermal Hall Conductivity}
\label{sec:transport}

We next formulate the intrinsic thermal Hall response carried by the bosonic
spinon bands.  Let \(j_\mu^{Q}\) denote the heat-current density in the
direction \(\mu=x,y\).  Its linear response to a slowly varying
temperature field is defined by
\begin{equation}
 j_\mu^{Q}=-\sum_{\nu=x,y}\kappa_{\mu\nu}
 \bigl(\bm{\nabla}T\bigr)_\nu.
 \label{eq:thermal_transport_definition}
\end{equation}
In particular, our sign convention for the transverse coefficient is
\(j_x^{Q}=-\kappa_{xy}\bigl(\bm{\nabla}T\bigr)_y\).  For the insulating spin
system studied in this paper, the heat current is the energy current obtained from the continuity
equation for the quadratic mean-field Hamiltonian.  The corresponding
transport coefficient is the energy-current Kubo response supplemented by the
energy-magnetization correction
\cite{Luttinger-1964,Qin-Niu-Shi-2011,Matsumoto-Murakami-2011,
Matsumoto-Murakami-2011-2}.  With the units fixed in
Sec.~\ref{sec:model}, we denote this intrinsic thermal Hall conductivity simply
by \(\kappa_{xy}\).
The positive-norm particle bands satisfy \(\sigma_{3,n}=+1\) and
\((\mathcal E_{\bk})_{nn}=\varepsilon_{n\bk}>0\).  For
Cartesian momentum components \(k_x\) and \(k_y\), we use the shorthand
\(\partial_x\equiv\partial/\partial k_x\) and
\(\partial_y\equiv\partial/\partial k_y\).  For
each such band \(n\), the interband curvature entering the bosonic Kubo
formula is
\begin{equation}
 \Omega^{xy}_{n\bk}=-2\,\mathrm{Im}\!\sum_{m\ne n}
 \sigma_{3,n}\sigma_{3,m}
 \frac{(T_{\bk}^{\dagger}\partial_x\mathcal M_{\bk}T_{\bk})_{nm}
       (T_{\bk}^{\dagger}\partial_y\mathcal M_{\bk}T_{\bk})_{mn}}
      {\bigl[(\sigma_3\mathcal E_{\bk})_{nn}
       -(\sigma_3\mathcal E_{\bk})_{mm}\bigr]^2}.
 \label{eq:kubo_curvature}
\end{equation}
Here the sum runs over both positive- and negative-norm intermediate states.
In particular, a hole intermediate state has \(\sigma_{3,m}=-1\), so the
denominator remains the difference of the signed dynamical-matrix
eigenvalues, exactly as in the Colpa representation.
The thermal Hall conductivity divided by temperature, \(\kappa_{xy}/T\), is
evaluated by summing over the magnetic Brillouin zone.  Explicitly,
\begin{align}
 \frac{\kt}{T}={}&-\frac{1}{A_{\mathrm M}N_k^2}
 \sum_{\bk\in\mathrm{MBZ}}\sum_n
 \left\{c_2[n_{\mathrm B}(\varepsilon_{n\bk})]
 -\frac{\pi^2}{3}\right\}\Omega^{xy}_{n\bk},
 \label{eq:kappa_kubo}\\
 c_2(x)={}&(1+x)\left(\ln\frac{1+x}{x}\right)^2
 -(\ln x)^2-2\operatorname{Li}_2(-x).
 \label{eq:c2}
\end{align}
The band sum in Eq.~\eqref{eq:kappa_kubo} runs over the positive-norm particle
bands.  Here \(A_{\mathrm M}\) is the real-space magnetic-unit-cell area.
With \(a=1\), it is \(3\sqrt{3}/2\), \(3\sqrt{3}\), and \(6\sqrt{3}\) for
the \(0\)-, \(\pi\)-, and \(\pi/2\)-flux cells, respectively.  The quantity
\(N_k\times N_k\) is the transport mesh in the magnetic Brillouin zone.
Because \(\Omega^{xy}_{n\bk}\) is defined using Cartesian momenta, the
discrete MBZ measure is \((A_{\mathrm M}N_k^2)^{-1}\).  Here
\(n_{\mathrm B}(\varepsilon)=[\exp(\varepsilon/T)-1]^{-1}\).  This is the
bosonic linear-response expression including the
energy-magnetization correction
\cite{Qin-Niu-Shi-2011,Katsura-Nagaosa-Lee-2010,Matsumoto-Murakami-2011,
Matsumoto-Murakami-2011-2,Shindou-Matsumoto-Murakami-Ohe-2013,
Mook-Henk-Mertig-2014,Kondo-Akagi-Katsura-2020,
Fujiwara-Kitamura-Morimoto-2022,Koyama-Nasu-2021,
Samajdar-Chatterjee-2019}.
Below, we refer to \(\kappa_{xy}/T\) as the reduced thermal Hall response.  All
transport curves and numerical response values use this quantity.
\label{para:physical_units_start}
\red{Equation~\eqref{eq:kappa_kubo} describes a single honeycomb layer.
Restoring physical units, the plotted values of \(\kappa_{xy}/T\) are
multiplied by \(k_{\mathrm B}^{2}/\hbar\), giving \(\mathrm{W\,K^{-2}}\).
Temperatures in kelvin follow by multiplying the plotted \(T/J_1\) by the
physical \(J_1/k_{\mathrm B}\).}
\label{para:physical_units_end}

We discretize the magnetic Brillouin zone by a uniform midpoint mesh and use
analytic derivatives
\(\partial_x\mathcal M_{\bk}\) and
\(\partial_y\mathcal M_{\bk}\).
Both positive--negative and positive--positive intermediate states are
retained.  The two orientations of each positive--positive pair are
combined into the stable difference
\(c_2[n_{\mathrm B}(\varepsilon_n)]-c_2[n_{\mathrm B}(\varepsilon_m)]\), so
the constant \(-\pi^2/3\) cancels within such a pair.  Internal pairs of an
exactly degenerate positive-energy multiplet have identical thermal weights
and are removed before division by their energy splitting.  This pair-combined
evaluation is algebraically identical to Eqs.~\eqref{eq:kubo_curvature} and
\eqref{eq:kappa_kubo} away from exact degeneracies and is invariant
under a unitary rotation within a degenerate multiplet.

\section{Results}
\label{sec:results}

\begin{figure}[t]
\centering
\includegraphics[width=\columnwidth]{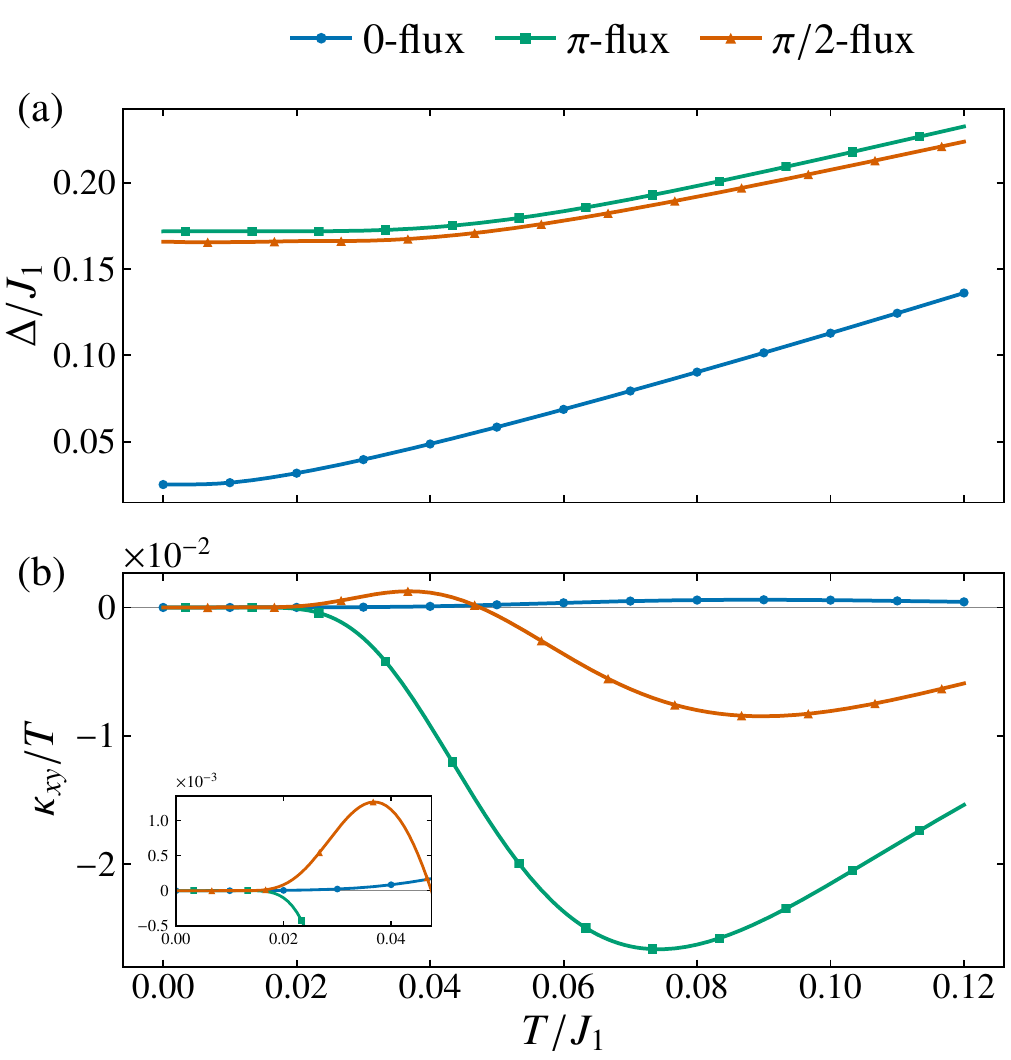}
\caption{
Direct comparison of the three flux sectors at the common parameters
\(J_2/J_1=0.20\), \(D/J_1=0.10\), and \(h/J_1=0.05\).
(a) Self-consistent spinon gaps.  (b) Intrinsic thermal Hall responses
\(\kappa_{xy}/T\).
The inset enlarges the low-temperature response near zero for all three
sectors.
}
\label{fig:common_flux_transport}
\end{figure}

\begin{figure*}[t]
\centering
\includegraphics[width=0.94\textwidth]{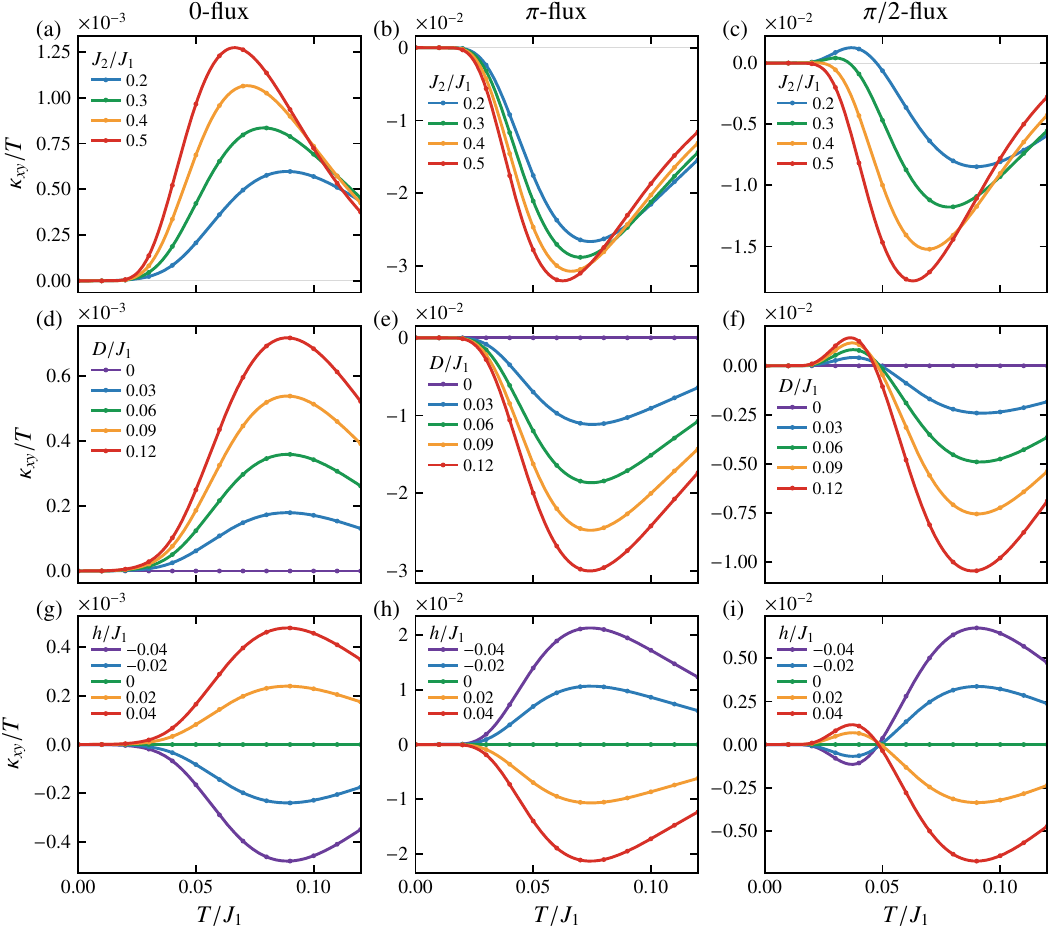}
\caption{
Parameter dependence of \(\kappa_{xy}/T\).  The columns show the \(0\)-,
\(\pi\)-, and \(\pi/2\)-flux branches, respectively.
(a)--(c) \(J_2/J_1=0.20\), \(0.30\), \(0.40\), and \(0.50\) at fixed
\(D/J_1=0.10\) and \(h/J_1=0.05\).
(d)--(f) \(D/J_1=0,0.03,0.06,0.09\), and \(0.12\) at fixed
\(J_2/J_1=0.20\) and \(h/J_1=0.05\).
(g)--(i) \(h/J_1=-0.04,-0.02,0,0.02\), and \(0.04\) at fixed
\(J_2/J_1=0.20\) and \(D/J_1=0.10\).
}
\label{fig:parameter_dependence}
\end{figure*}

\begin{figure}[t]
\centering
\includegraphics[width=\columnwidth]{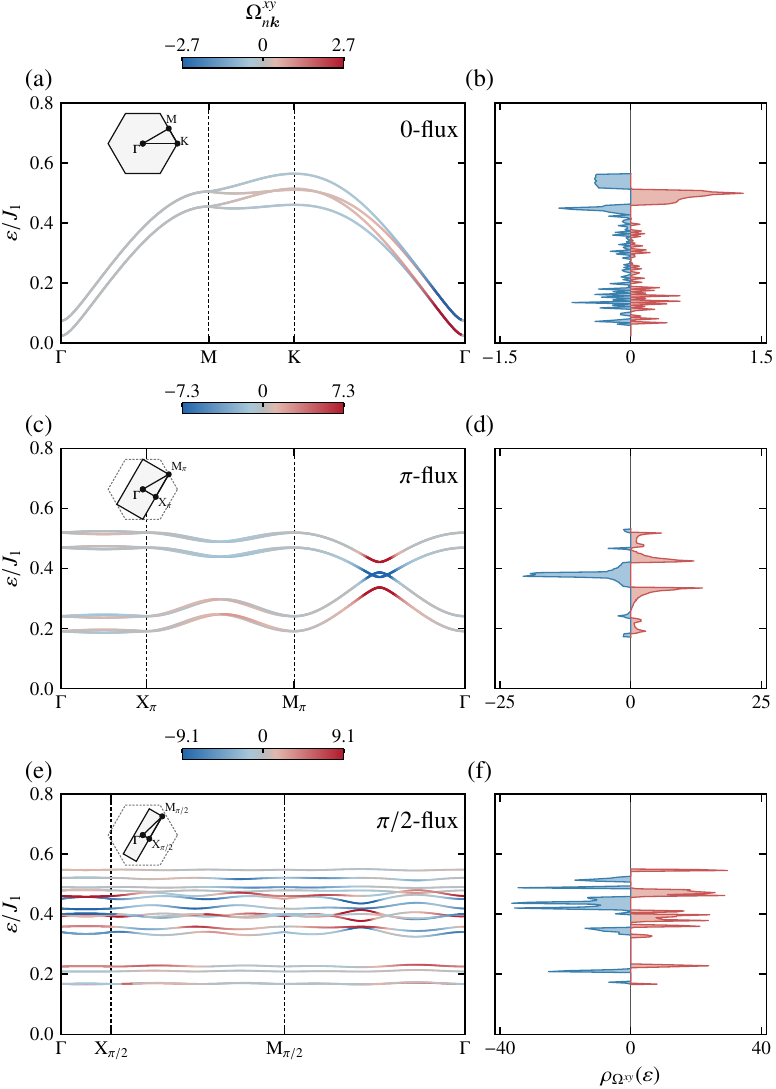}
\caption{
Positive-energy bosonic BdG bands and Berry-curvature distributions at
\(T=0\) for the common parameters \(J_2/J_1=0.20\), \(D/J_1=0.10\), and
\(h/J_1=0.05\).  Rows show the \(0\)-, \(\pi\)-, and \(\pi/2\)-flux sectors.
The left panels show bands colored by their Berry curvature, with
the paths shown in the insets.  The right panels show the energy-resolved
signed Berry-curvature density.  The Berry-curvature color scale is chosen
independently for each row.
}
\label{fig:band_berry_curvature}
\end{figure}

\label{para:constraint_comparison_start}
\red{The main three-sector comparison uses \(\kappa=\sqrt{3}-1\).  At the common
microscopic parameters, the \(0\)-flux branch condenses at zero temperature
when \(\kappa=1\) is used, so the comparison of three gapped, uncondensed
branches relies on the corrected prescription.  Appendix~\ref{app:kappa_one}
provides a separate check at \(\kappa=1\) for the two branches that remain
gapped.}
\label{para:constraint_comparison_end}
All thermal Hall values quoted below refer to \(\kappa_{xy}/T\).  Unless a
field or DM sign is explicitly varied, the reported signs use \(D>0\) and
\(h>0\), with the DM-arrow convention of Fig.~\ref{fig:model_bz}(a) and the
thermal-current convention of Eq.~\eqref{eq:thermal_transport_definition}.  The
self-consistent mean fields are evaluated on a \(60\times60\) mesh, while
the Berry curvature and \(\kappa_{xy}/T\) are evaluated on a
\(121\times121\) mesh.  We verified convergence by increasing the respective
mesh sizes to \(120\times120\) and \(201\times201\).

Figure~\ref{fig:common_flux_transport}(a) shows that all three branches
remain gapped throughout the plotted temperature interval.  Their gaps
increase with temperature without changing their ordering.  The \(0\)-flux
branch has the smallest gap, while the \(\pi\)-flux branch has the largest.
In Fig.~\ref{fig:common_flux_transport}(b),
the \(0\)-flux response, although small on this scale, is positive, whereas the
\(\pi\)-flux response is negative over the resolved finite-response range.
The \(\pi/2\)-flux response is positive at low temperature, crosses zero at
\(T/J_1\simeq0.0475\), and is negative
above the crossing.

Across \(0.20\leq J_2/J_1\leq0.50\), Figs.~\ref{fig:parameter_dependence}(a)
and \ref{fig:parameter_dependence}(b) show a positive \(0\)-flux response and
a dominant negative \(\pi\)-flux response.  In panel (c), the positive
low-temperature window of the \(\pi/2\)-flux response shrinks as \(J_2\)
increases.  Its zero crossing moves from \(T/J_1\simeq0.0475\) at
\(J_2/J_1=0.20\) to \(0.0242\) at \(J_2/J_1=0.40\), and the
\(J_2/J_1=0.50\) curve is nonpositive on the plotted grid.  In panels
(d)--(f), all three responses vanish at \(D=0\).  Increasing \(D\) enhances
their magnitudes without changing these qualitative trends over the range
shown.  Panels (g)--(i) show that the responses vanish at \(h=0\), are odd
under \(h\to-h\), and increase in magnitude with \(|h|\).
\label{para:j2_window_start}
\red{Table~S2 of the Supplemental Material~\cite{SupplementalMaterial}
lists the verified uncondensed ranges for all three sectors, and Fig.~S4
summarizes the \(\pi/2\)-flux crossing temperature and positive
low-temperature peak versus \(J_2/J_1\).  The \(\pi/2\)-flux response
exhibits a sign reversal at the sampled points from \(J_2/J_1=0.20\)
to \(0.4656\).}
\label{para:j2_window_end}

To compare the band geometry with the transport response, we define the
energy-resolved Berry-curvature density by
\begin{equation}
 \rho_{\Omega^{xy}}(\varepsilon)
 =\frac{1}{A_{\mathrm M}N_k^2}
 \sum_{\bk\in\mathrm{MBZ}}\sum_n
 \Omega^{xy}_{n\bk}\,
 \delta\!\left(\varepsilon-\varepsilon_{n\bk}\right).
 \label{eq:berry_curvature_density}
\end{equation}
The band sum again runs over positive-norm particle bands.  For
Fig.~\ref{fig:band_berry_curvature}, the two U($1$) blocks are included in the
sum.

Figure~\ref{fig:band_berry_curvature} shows the \(T=0\) positive-energy bands
and their Berry-curvature distributions for the same Hamiltonian parameters
in all three flux sectors.  The \(0\)-, \(\pi\)-, and \(\pi/2\)-flux BdG Hamiltonians
contain two, four, and eight positive bands per U($1$) block, respectively.
The \(0\)-flux bands are plotted along
\(\Gamma\)-\(\mathrm{M}\)-\(\mathrm{K}\)-\(\Gamma\), while the finite-flux bands are plotted
along \(\Gamma\)-\(\mathrm{X}_{\pi}\)-\(\mathrm{M}_{\pi}\)-\(\Gamma\) and
\(\Gamma\)-\(\mathrm{X}_{\pi/2}\)-\(\mathrm{M}_{\pi/2}\)-\(\Gamma\), respectively,
in the corresponding magnetic Brillouin zones.  Along the displayed path in panel (a), the largest curvature
magnitudes occur on the
nearly degenerate low-energy bands close to \(\Gamma\) along the
\(\mathrm K\)-to-\(\Gamma\) segment.  In panel (c), strong opposite-sign
weights occur on adjacent bands along the \(\mathrm M_{\pi}\)-to-\(\Gamma\)
segment.  The strongest opposite-sign pair in panel (e) also lies on the
\(\mathrm M_{\pi/2}\)-to-\(\Gamma\) segment, with weaker curvature distributed among
the other bands.  Panels (b),
(d), and (f) show both positive and negative contributions to the
energy-resolved signed density.  Compared with the \(0\)-flux sector, the two
finite-flux sectors exhibit substantially larger signed-density features,
with both signs present over the plotted energy range.
\red{Complementary unweighted Berry-curvature maps on both sides of the
\(\pi/2\)-flux sign reversal are given in Fig.~S2 of the Supplemental
Material~\cite{SupplementalMaterial}.}

\section{Discussion}
\label{sec:discussion}

\begin{figure}[t]
\centering
\includegraphics[width=\columnwidth]{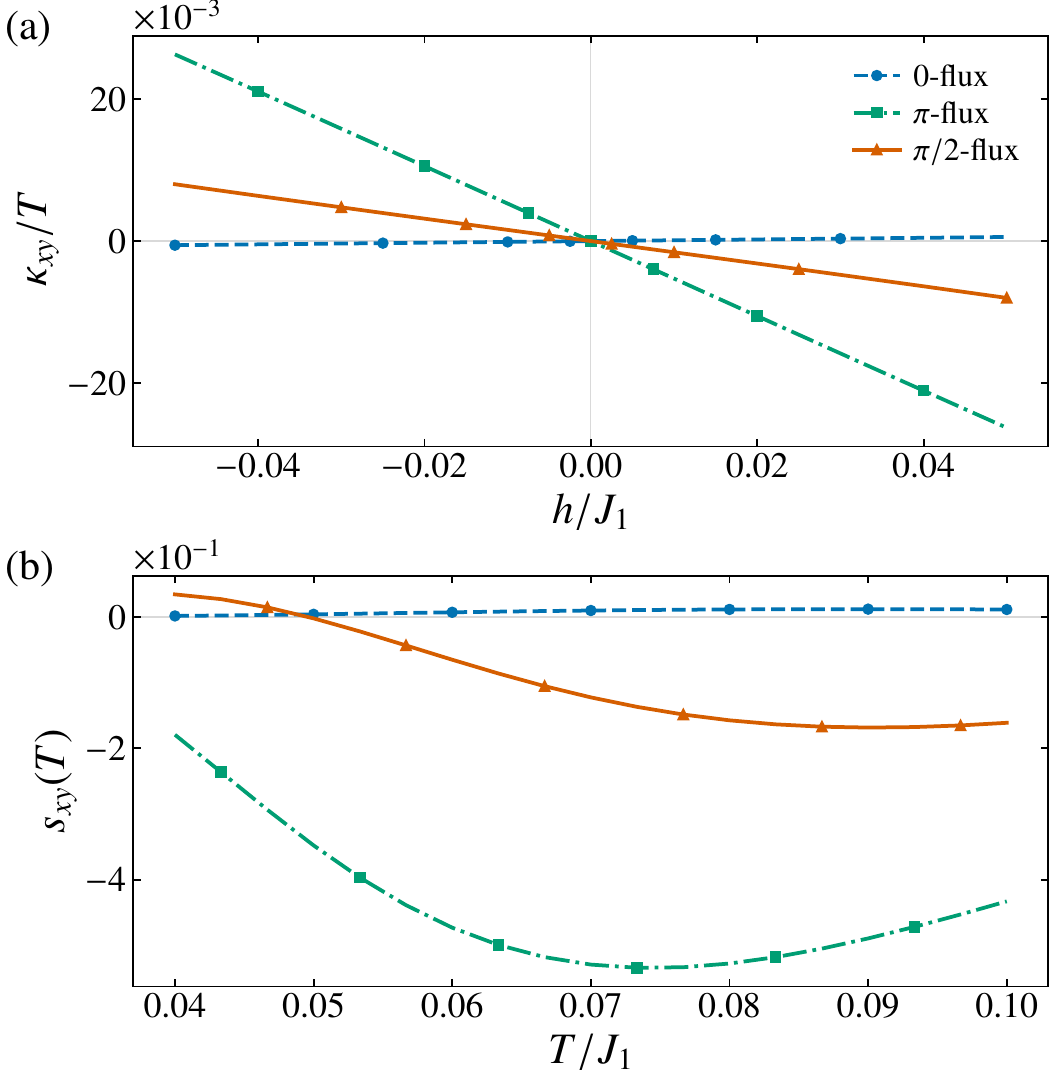}
\caption{
Small-field thermal Hall response at \(J_2/J_1=0.20\) and \(D/J_1=0.10\).
(a) Field dependence of \(\kappa_{xy}/T\) at \(T/J_1=0.08\).
(b) Local slope
\(s_{xy}(T)=
\left.\partial(\kt/T)/\partial(h/J_1)\right|_{h=0}\), obtained after
antisymmetrizing the \(\pm h\) data and fitting an odd cubic polynomial through
the origin.
}
\label{fig:small_field}
\end{figure}

\begin{figure}[t]
\centering
\includegraphics[width=\columnwidth]{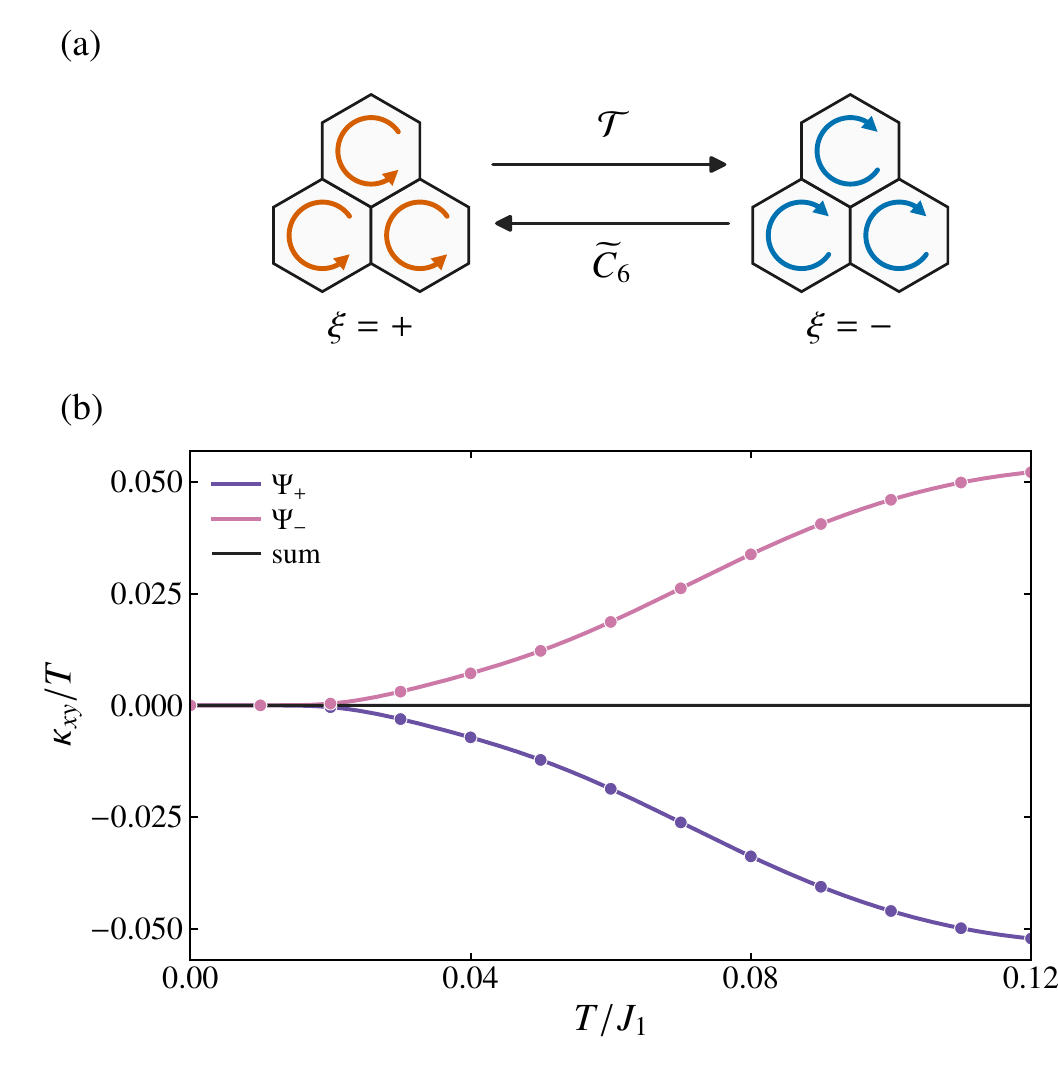}
\caption{
Zero-field cancellation in a fixed chiral \(\pi/2\)-flux domain.
(a) The orange and blue lattice patches denote the two lattice-wide chiral
domains \(\xi=+1\) and \(\xi=-1\).  The sketches represent the periodic
Wilson-loop pattern rather than physical bond currents.  Time reversal
\(\mathcal T\) and the
projective sixfold operation \(\widetilde C_6\) exchange these domains.
Within one selected domain, both
U($1$) BdG blocks \(\Psi_+\) and \(\Psi_-\) remain present and are related by
\(\widetilde C_6\mathcal T\) at \(h=0\).
(b) The two block-resolved contributions to \(\kappa_{xy}/T\) are finite and
opposite, while their physical sum vanishes.  Here \(J_2/J_1=0.20\) and
\(D/J_1=0.10\).
}
\label{fig:chiral_h0_cancellation}
\end{figure}

We first discuss the symmetries that constrain the thermal Hall response,
then relate its flux dependence to band geometry and clarify the scope of
the proposed transport diagnostic.
Chirality alone does not guarantee a finite thermal Hall response.
As illustrated in Fig.~\ref{fig:chiral_h0_cancellation}(a), time reversal
\(\mathcal T\) and the projective sixfold operation \(\widetilde C_6\)
each exchange the two conjugate chiral domains.  Their product
\(\widetilde C_6\mathcal T\) preserves either domain at \(h=0\).
Since a proper rotation leaves \(\kappa_{xy}\) unchanged while time reversal
reverses its sign, this antiunitary relation gives
\(\kappa_{xy}^{(\xi)}(h)=-\kappa_{xy}^{(\xi)}(-h)\) and hence
\(\kappa_{xy}^{(\xi)}(0)=0\) for each domain \(\xi=\pm1\).
At zero field, the same operation pairs states in the two U($1$) spin blocks
with equal energies and opposite Berry curvatures at symmetry-related momenta.
The blocks \(\Psi_+\) and \(\Psi_-\) coexist within a selected domain;
they are not domain labels.  Their equal Bose weights therefore make the
two finite block-resolved Hall responses cancel, as shown in
Fig.~\ref{fig:chiral_h0_cancellation}(b).  A finite field breaks the internal
symmetry and allows a net response, while retaining the field-reversal
relation in Fig.~\ref{fig:small_field}(a).  The block definitions and symmetry
transformations are given in Appendix~\ref{app:chiral_zero_field}.

The vanishing response at \(D=0\) has a distinct antiunitary origin.  For the
\(0\)- and \(\pi\)-flux sectors, \(\mathcal R_x(\pi)\mathcal T\), where
\(\mathcal R_x(\pi)\) is a spin rotation by \(\pi\) about \(\hat{\bm x}\),
preserves \(h\) while mapping \(D\) to \(-D\).  In a fixed chiral
\(\pi/2\)-flux domain, \(\mathcal T\) also exchanges the two domains, and the
projective operation \(\widetilde C_6\) returns the transformed Ansatz to the
original domain.  Thus the relevant operation is
\(\widetilde C_6\mathcal R_x(\pi)\mathcal T\).  Both relations reverse the
Hall response and imply
\begin{equation}
 \kappa_{xy}(D,h)=-\kappa_{xy}(-D,h),
 \qquad
 \kappa_{xy}(0,h)=0.
 \label{eq:discussion_dm_odd}
\end{equation}
Together with oddness in \(h\), these constraints make \(Dh\) the
leading symmetry-allowed dependence near \(D=h=0\) for an analytic gapped
branch.  The zeros in
Figs.~\ref{fig:parameter_dependence}(d)--\ref{fig:parameter_dependence}(f)
are therefore symmetry enforced rather than accidental.  The transformation
of the Hamiltonian is given in Appendix~\ref{app:chiral_zero_field}.

At \(T/J_1=0.08\), Fig.~\ref{fig:small_field}(a) shows a small
positive \(0\)-flux response and negative \(\pi\)- and \(\pi/2\)-flux
responses for \(h>0\).  The finite-flux responses are also more sensitive
to the field than the \(0\)-flux response.  Figure~\ref{fig:small_field}(b)
shows the fitted low-field coefficient \(s_{xy}(T)\).  The
\(0\)-flux coefficient is positive and increases throughout the plotted
interval.  The \(\pi\)-flux coefficient is negative, reaches its largest
magnitude around \(T/J_1\simeq0.09\), and then decreases slightly in
magnitude.  The \(\pi/2\)-flux coefficient is positive at the lowest plotted
temperatures, crosses zero near \(T/J_1\simeq0.050\), and becomes negative as
the temperature is raised further.  Its magnitude remains smaller than that
of the \(\pi\)-flux coefficient over the displayed range.

Appendix~\ref{app:similar_gap} shows that matching the lowest spinon gap does
not collapse the three thermal Hall curves.  The band analysis in
Fig.~\ref{fig:band_berry_curvature} demonstrates that the sectors also differ
in band geometry.  Within the quadratic spinon
description, the thermal Hall response probes the positive-energy BdG
eigenvectors, their interband matrix elements, and the thermal occupations
simultaneously.  Magnetic folding changes both the number of positive bands per
U($1$) block and the distribution of Berry curvature over the magnetic
Brillouin zone.  The response in Eq.~\eqref{eq:kappa_kubo} is therefore not a
sum of bare Berry curvatures alone.  It also includes Bose factors and both
positive--negative and positive--positive interband processes.  Consequently,
an integer Chern number of a bosonic BdG band, even when it is well defined for
an isolated band, does not imply a quantized thermal Hall plateau for the
uncondensed spinon gas
\cite{Mook-Henk-Mertig-2014,McClarty-2022,Zhang-Gao-Chen-2024}.

The temperature-driven sign reversal of the \(\pi/2\)-flux branch can be
resolved directly in energy.  Figure~S3 of the Supplemental
Material~\cite{SupplementalMaterial} shows the thermal-kernel-weighted
Berry-curvature density on the two sides of the crossing.  Using the
self-consistent BdG Hamiltonian at each temperature, its energy integral is
positive at \(T/J_1=0.03\) and negative at \(T/J_1=0.08\).  By contrast, when
the \(T=0\) BdG Hamiltonian is held fixed and only the Bose kernel is varied,
the response remains positive throughout the crossing region.  The sign
reversal therefore requires the temperature dependence of the self-consistent
BdG Hamiltonian, not only the Bose occupation factor.

\label{para:ordered_transport_start}
\red{Magnetic order can also produce a finite thermal Hall response, as shown
for field-canted N\'eel magnons with next-nearest-neighbor DM
coupling~\cite{Owerre-2017-Neel-Hall}.  Our supplementary linear-spin-wave
benchmarks at \(J_2/J_1=0,0.05,0.10\), \(D/J_1=0.10\), and
\(h/J_1=0.05,0.20\) give positive \(\kappa_{xy}/T\), with no resolved
sign reversal over \(0<T/J_1\leq0.12\).
These low-\(J_2\) results are not a comparison at the common
\(J_2/J_1=0.20\) point, where the tested spin-wave references do not
provide a reliable harmonic description.  The calculations and applicability checks are
given in the Supplemental Material~\cite{SupplementalMaterial}.
A finite thermal Hall response alone therefore does not establish
fractionalization.}
\label{para:ordered_transport_end}

\label{para:dsf_scope_start}
The dynamical structure factor constrains the two-spinon threshold, dispersion,
and spin-operator coherence factors
\cite{Knolle-2014,Messio-Cepas-2010}, whereas \(\kappa_{xy}\) depends on the
Berry-curvature distribution, interband matrix elements, and thermal
occupations across the spectrum.  \red{The dynamical structure factor thus
provides a complementary characterization of the candidate states but does
not enter the thermal Hall calculation.}  The corresponding spectra and
equal-time correlations are given in the Supplemental
Material~\cite{SupplementalMaterial}.
\label{para:dsf_scope_end}

\Needspace{3\baselineskip}
\label{para:branch_interpretation_start}
\red{Self-consistency within a fixed flux Ansatz does not establish that the
corresponding state is realized in equilibrium.  State selection depends on
microscopic interactions and competition with other spin-liquid and ordered
states.}
\label{para:valence_bond_scope_start}
\red{Our comparison also omits an independent plaquette valence-bond-order
Ansatz and its competition with the flux states.
The verified noncondensed ranges are therefore not equilibrium phase boundaries.
The resolved sign-change window includes, for example,
\(0.22<J_2/J_1<0.25\), where spin-\(1/2\) DMRG reports a nonmagnetic regime
consistent with a possible spin liquid~\cite{Gong-Sheng-Motrunich-Fisher-2013}.
This overlap alone does not establish a spin-liquid phase in the present
model.  The phase boundaries obtained at \(D=h=0\) cannot be assumed to
remain unchanged at finite \(D\) and \(h\).}
\label{para:valence_bond_scope_end}
\red{The calculated thermal Hall responses are therefore predictions for
candidate gapped spinon states.  When independent evidence supports such a
description, the Hall sign and temperature dependence can help distinguish
its flux structure, complementing the excitation spectrum and dynamical
structure factor.}
\label{para:branch_interpretation_end}

At the common point in Fig.~\ref{fig:common_flux_transport}, for \(D>0\) and
\(h>0\), with the DM-arrow convention of Fig.~\ref{fig:model_bz}(a) and the
thermal-current convention of Eq.~\eqref{eq:thermal_transport_definition}, the branches
associated with the three flux sectors can be distinguished from the full
temperature dependence rather than from a single magnitude.  The \(0\)-flux
curve stays positive throughout the temperature range shown,
the \(\pi\)-flux curve is negative, and the
\(\pi/2\)-flux curve exhibits a pronounced sign reversal from positive
to negative.  This pronounced reversal distinguishes the two finite-flux
branches in a low-temperature interval in which the \(0\)- and \(\pi/2\)-flux
responses have the same sign.  Absolute magnitudes remain more sensitive to
the saddle-point parameters than this combined sign-and-temperature signature.

\label{para:constraint_robustness_start}
Appendix~\ref{app:kappa_one} provides a separate check of the boson-number
prescription.  When \(\kappa=1\) is used, the \(0\)-flux saddle-point solution
condenses \red{at zero temperature} and is therefore excluded from the
gapped-branch comparison for the same
Hamiltonian parameters.  The \(\pi\)- and \(\pi/2\)-flux saddle-point solutions
remain gapped,
however, and retain the same qualitative transport patterns.  The
\(\pi\)-flux response shown in Fig.~\ref{fig:kappa_one_appendix} is negative,
while the \(\pi/2\)-flux response changes
from positive to negative as temperature increases.  \red{Thus, the universal
aspect across the two constraint prescriptions examined here is the
qualitative temperature evolution of these gapped branches.  The response
magnitudes and characteristic temperature scales depend quantitatively on
the prescription.  The \(\pi/2\)-flux crossing shifts from
\(T/J_1\simeq0.0475\) at \(\kappa=\sqrt{3}-1\) to \(0.063\)
at \(\kappa=1\).}
\label{para:constraint_robustness_end}

The quadratic calculation omits gauge fluctuations, spinon interactions, and
phonons, all of which can renormalize or add to the intrinsic response
\cite{Read-Sachdev-1991,Sachdev-1992,Koyama-Nasu-2024,Akazawa-2020}.
Gauge-flux excitations may also carry heat or scatter spinons
\cite{Joy-Rosch-2022}.  In the ideal model, the two \(\pi/2\)-flux domains give
the same field-induced \(\kappa_{xy}\).  Their populations become relevant only
when additional perturbations break this relation.
\red{Interpreting the absolute Hall sign requires an independent constraint
on the DM sign relative to the crystal structure.  Its experimental
determination without long-range order remains challenging.
Polarized-neutron diffraction exploits ordered moments~\cite{Thoma-2021},
whereas microscopic spin-orbit calculations provide a material-specific
theoretical constraint~\cite{Heully-Alary-2025}.}

\label{para:higher_spin_outlook_start}
\red{The present results concern spin-\(1/2\).  The same transport framework can be applied at larger
\(S\), including \(S=3/2\), only while the self-consistent saddle remains
gapped and uncondensed under the corresponding boson-number prescription.
For spin-\(1\), DMRG of the \(J_1-J_2\) model
without DM coupling or a field finds evidence for plaquette valence-bond
order in a narrow nonmagnetic region between magnetically ordered
phases~\cite{Gong-Zhu-Sheng-2015}.  The stability of candidate spin-liquid
states against magnetic and valence-bond order would need to be assessed
separately in such an extension.  Ordered phases require a consistent
treatment of the condensate and collective modes beyond the present saddle-point
approximation~\cite{Zhang-Ghioldi-2019,Zhang-Ghioldi-2022}.
Extending this description to thermal Hall transport remains a task for
future work.}
\label{para:higher_spin_outlook_end}

\section{Summary}
\label{sec:summary}

In this work, we constructed self-consistent \(0\)-, \(\pi\)-, and
\(\pi/2\)-flux Schwinger-boson mean-field branches for the honeycomb
\(J_1-J_2\) model with a next-nearest-neighbor Dzyaloshinskii--Moriya
interaction and a magnetic field, and evaluated their intrinsic thermal Hall
conductivity from the Berry curvature of the bosonic Bogoliubov bands.  By comparing the
three sectors at identical
\(J_2/J_1\), \(D/J_1\), and \(h/J_1\), rather than ranking them by their
mean-field free energies, we directly identified how their distinct flux
structures affect thermal transport.

At the common parameter point, the \(0\)- and \(\pi\)-flux responses have
opposite signs in the low-temperature window examined here, whereas the
\(\pi/2\)-flux response changes sign as the temperature increases.  In a
separate gap-matched control in Appendix~\ref{app:similar_gap}, \(J_2/J_1\) is
adjusted independently for each sector, yet the thermal Hall curves remain
distinct.  These differences originate not
from the gap alone but from flux-dependent bosonic BdG eigenvectors,
interband matrix elements, and the energy distribution
of Berry curvature.  The sign and temperature dependence of the response thus
provide a flux-sensitive transport signature within the mean-field regime
studied.  We further showed that a projective antiunitary symmetry forces the
thermal Hall conductivity of a fixed chiral \(\pi/2\)-flux domain to vanish at
zero field despite its broken time-reversal symmetry.  More generally, the
response is odd separately in \(D\) and \(h\); reversing either one reverses
\(\kappa_{xy}\), with \(\kappa_{xy}\propto Dh\) near the origin.

The present calculation evaluates the intrinsic thermal Hall response arising
from the Berry curvature and band geometry of uncondensed bosonic spinons
within SBMFT.
Quantitative comparison with materials will require effects beyond quadratic
SBMFT, including spinon interactions, gauge fluctuations, and phonons.

\begin{acknowledgments}
The author thanks J.~Nasu and R.~Mukherjee for helpful discussions and
comments.
The author also thanks A.~Ralko and J.~Merino for valuable discussions of the
three honeycomb-lattice flux Ansätze considered in this work.
Parts of the numerical calculations were performed on the supercomputing
systems at ISSP, the University of Tokyo.
The author acknowledges support from GP-Spin at Tohoku University.
\end{acknowledgments}

\appendix
\section{Antiunitary Constraints at Zero Field and Zero DM Coupling}
\label{app:chiral_zero_field}

This appendix gives the block-level symmetry argument underlying
Fig.~\ref{fig:chiral_h0_cancellation}.  We distinguish the chiral-domain label
\(\xi=\pm1\) from the BdG-block label \(s=\pm\).  The former denotes the two
conjugate, lattice-wide \(\pi/2\)-flux domains.  If the gauge-invariant singlet
Wilson loop defined in Eq.~\eqref{eq:singlet_wilson_loop} is written as
\begin{equation}
 W_p^{(\xi)}=
 \lvert W_p\rvert e^{i\xi\pi/2},
 \qquad \xi=\pm1,
 \label{eq:appendix_chiral_wilson_loop}
\end{equation}
then complex conjugation interchanges the two domains.  By contrast, the
labels \(+\) and \(-\) on the Nambu spinors below identify two U($1$)-spin BdG
blocks that coexist within either fixed domain.  They are not domain labels.
The arrows drawn inside the plaquettes in
Fig.~\ref{fig:chiral_h0_cancellation}(a) are therefore only a schematic of the
conjugate Wilson-loop phases in Eq.~\eqref{eq:appendix_chiral_wilson_loop}, not
physical bond currents.

The out-of-plane DM interaction preserves spin rotations about
\(\hat{\bm z}\), and the resulting U($1$)-symmetric mean-field Hamiltonian can
consequently be reordered into two independent bosonic BdG blocks.  For a magnetic cell with
\(M\) sites, we define
\begin{align}
 \Psi_{\bk,+}^{\dagger}
 &\equiv
 \bigl(
 b_{\bk 1\uparrow}^{\dagger},\ldots,
 b_{\bk M\uparrow}^{\dagger},
 b_{-\bk 1\downarrow},\ldots,
 b_{-\bk M\downarrow}
 \bigr),
 \label{eq:appendix_psi_plus}\\
 \Psi_{\bk,-}^{\dagger}
 &\equiv
 \bigl(
 b_{\bk 1\downarrow}^{\dagger},\ldots,
 b_{\bk M\downarrow}^{\dagger},
 b_{-\bk 1\uparrow},\ldots,
 b_{-\bk M\uparrow}
 \bigr).
 \label{eq:appendix_psi_minus}
\end{align}
Thus, after a permutation of the components in
Eq.~\eqref{eq:nambu_spinor}, the full spinor is
\(\Psi_{\bk,+}\oplus\Psi_{\bk,-}\), and the quadratic Hamiltonian in a fixed
domain takes the form
\begin{equation}
 \mathcal H_{\mathrm{MF}}^{(\xi)}(h)
 =\frac12\sum_{\bk,s=\pm}
 \Psi_{\bk,s}^{\dagger}
 \mathcal M_{\bk,s}^{(\xi)}(h)
 \Psi_{\bk,s}.
 \label{eq:appendix_block_hamiltonian}
\end{equation}
The corresponding Hermitian blocks are
\begin{align}
 \mathcal M_{\bk,+}^{(\xi)}(h)
 &=
 \begin{pmatrix}
  H_{\uparrow}^{(\xi)}(\bk,h)
  &\Delta_{\uparrow\downarrow}^{(\xi)}(\bk)\\
  \Delta_{\uparrow\downarrow}^{(\xi)\dagger}(\bk)
  &H_{\downarrow}^{(\xi)T}(-\bk,h)
 \end{pmatrix},
 \label{eq:appendix_m_plus}\\
 \mathcal M_{\bk,-}^{(\xi)}(h)
 &=
 \begin{pmatrix}
  H_{\downarrow}^{(\xi)}(\bk,h)
  &\Delta_{\downarrow\uparrow}^{(\xi)}(\bk)\\
  \Delta_{\downarrow\uparrow}^{(\xi)\dagger}(\bk)
  &H_{\uparrow}^{(\xi)T}(-\bk,h)
 \end{pmatrix}.
 \label{eq:appendix_m_minus}
\end{align}
Here \(H_{\mu}^{(\xi)}\) contains the constraint, normal bond fields, DM
coupling, and the Zeeman shift for spin \(\mu\), whereas
\(\Delta_{\uparrow\downarrow}^{(\xi)}\) and
\(\Delta_{\downarrow\uparrow}^{(\xi)}\) contain the anomalous pairing terms.
For the saddle-point manifold used in the reported calculations, these terms
are generated by the nearest-neighbor singlet field \(\mathcal A_1\).  The
formal U($1$)-symmetric bond basis also contains the spin-vector pairing
operator \(\mathcal D^z\), but \(D_1^z\) and \(D_2^z\) are not independent
saddle coordinates and are not fed back into the BdG Hamiltonian, as specified
in the Supplemental Material~\cite{SupplementalMaterial}.
The physical response is obtained by summing the contributions from the two
blocks.  Each block has the Nambu metric
\begin{equation}
 \tau_3=\begin{pmatrix}\bm{1}_M&0\\0&-\bm{1}_M\end{pmatrix},
 \qquad
 \tau_3\mathcal M_{\bk,s}^{(\xi)}
 \lvert u_{n\bk,s}^{(\xi)}\rangle
 =\varepsilon_{n\bk,s}^{(\xi)}
 \lvert u_{n\bk,s}^{(\xi)}\rangle,
 \label{eq:appendix_block_colpa}
\end{equation}
where $n$ runs only over the positive-norm particle modes used below, with
$\varepsilon_{n\bk,s}^{(\xi)}>0$.

The antiunitary time-reversal operation acts on the Schwinger bosons as
\begin{equation}
 \mathcal T b_{i\uparrow}\mathcal T^{-1}=b_{i\downarrow},
 \qquad
 \mathcal T b_{i\downarrow}\mathcal T^{-1}=-b_{i\uparrow},
 \qquad
 \mathcal T i\mathcal T^{-1}=-i.
 \label{eq:appendix_time_reversal_bosons}
\end{equation}
It complex conjugates the Wilson loop, reverses the Zeeman field, and hence
maps the complete mean-field Hamiltonian according to
\begin{equation}
 \mathcal T\mathcal H_{\mathrm{MF}}^{(\xi)}(h)\mathcal T^{-1}
 =\mathcal H_{\mathrm{MF}}^{(-\xi)}(-h).
 \label{eq:appendix_time_reversal_hamiltonian}
\end{equation}
The sixfold lattice operation must be implemented projectively on spinons.
Writing \(p_6(m)\) for the site permutation induced by the rotation, and
\(\bm d_{6m}\) for the magnetic-cell translation needed to return the rotated
site to the chosen cell, its action can be represented as
\begin{equation}
 \widetilde C_6 b_{\bk m\mu}\widetilde C_6^{-1}
 =e^{i\vartheta_m^{(6)}}
  e^{-i(R_6\bk)\cdot\bm d_{6m}}
  b_{R_6\bk,p_6(m),\mu}.
 \label{eq:appendix_projective_c6}
\end{equation}
The compensating PSG phases \(e^{i\vartheta_m^{(6)}}\) depend on the
eight-site magnetic-cell convention.  Their explicit values are not needed
to determine how \(\widetilde C_6\) acts on the gauge-invariant Wilson loop.
Here \(C_6\) denotes a proper rotation by \(\pi/3\), which
advances the six sites by one position around the
hexagon and exchanges the A and B sublattices.  Consequently the alternating
conjugation pattern is shifted by one bond according to
\begin{align}
 C_6\,W_p\longmapsto{}&
 \mean{\mathcal A_{i_2i_3}}
 \left(-\mean{\mathcal A_{i_3i_4}}^{*}\right)
 \mean{\mathcal A_{i_4i_5}}
 \notag\\[-2pt]
 &\times
 \left(-\mean{\mathcal A_{i_5i_6}}^{*}\right)
 \mean{\mathcal A_{i_6i_1}}
 \left(-\mean{\mathcal A_{i_1i_2}}^{*}\right)
 \notag\\
 ={}&W_p^*.
 \label{eq:appendix_c6_wilson}
\end{align}
This conjugation is not a reversal of a geometrical circulation.  A proper
rotation preserves clockwise versus counterclockwise orientation.  The conjugation instead
comes from the sublattice exchange in the alternating pairing-field Wilson
loop.  Since \(W_p\) is gauge invariant, the compensating PSG phases cancel
from Eq.~\eqref{eq:appendix_c6_wilson}.  Thus, for the representative used
here, \(\widetilde C_6\) obeys
\begin{equation}
 \widetilde C_6\mathcal H_{\mathrm{MF}}^{(\xi)}(h)
 \widetilde C_6^{-1}
 =\mathcal H_{\mathrm{MF}}^{(-\xi)}(h).
 \label{eq:appendix_c6_hamiltonian}
\end{equation}
Consequently a fixed chiral domain is closed under
\(\widetilde C_3=\widetilde C_6^2\), not under \(\widetilde C_6\) itself.
The self-consistent bond and onsite fields in the eight-site cell
respect this retained symmetry.

We now restore \(D\) as an explicit Hamiltonian argument and consider the
spin rotation \(\mathcal R_x(\pi)\).  The antiunitary product
\(\mathcal X\equiv\mathcal R_x(\pi)\mathcal T\) acts on the spin components as
\begin{equation}
 \mathcal X(S_i^x,S_i^y,S_i^z)\mathcal X^{-1}
 =(-S_i^x,S_i^y,S_i^z).
 \label{eq:appendix_RxT_spin}
\end{equation}
It leaves both Heisenberg terms and the Zeeman term invariant, while
\(\hat{\bm z}\cdot(\bm S_i\times\bm S_j)\) changes sign.  Since the 0- and
\(\pi\)-flux representatives are invariant under time reversal up to their
PSG gauge transformations, their complete mean-field Hamiltonians obey
\begin{equation}
 \mathcal X\mathcal H_{\mathrm{MF}}^{(\phi)}(D,h)\mathcal X^{-1}
 =\mathcal H_{\mathrm{MF}}^{(\phi)}(-D,h),
 \qquad \phi=0,\pi.
 \label{eq:appendix_dm_reversal_zero_pi}
\end{equation}
For a fixed \(\pi/2\)-flux domain, \(\mathcal X\) exchanges \(\xi\) with
\(-\xi\).  Equation~\eqref{eq:appendix_c6_hamiltonian} then shows that
\(\Xi\equiv\widetilde C_6\mathcal X\) returns the transformed Ansatz to the
same domain.  This gives
\begin{equation}
 \Xi\mathcal H_{\mathrm{MF}}^{(\xi)}(D,h)\Xi^{-1}
 =\mathcal H_{\mathrm{MF}}^{(\xi)}(-D,h).
 \label{eq:appendix_dm_reversal_pi2}
\end{equation}
The spatial part of either operation is absent or a proper rotation.
Consequently, the antiunitary transformation preserves the energy at the
related momentum and reverses the Berry curvature.  After summing
over the magnetic Brillouin zone, Eqs.~\eqref{eq:appendix_dm_reversal_zero_pi}
and \eqref{eq:appendix_dm_reversal_pi2} give, in all three sectors,
\begin{equation}
 \kappa_{xy}(D,h)=-\kappa_{xy}(-D,h),
 \qquad \kappa_{xy}(0,h)=0.
 \label{eq:appendix_kappa_dm_odd}
\end{equation}
This transformation was also checked directly for the gauge representatives
used in the numerical calculation.  The spectra at \(D\) and
\(-D\) coincide, whereas their integrated Hall responses have opposite signs.

The product
\(\Theta\equiv\widetilde C_6\mathcal T\) is antiunitary.  Combining
Eqs.~\eqref{eq:appendix_time_reversal_hamiltonian} and
\eqref{eq:appendix_c6_hamiltonian} gives
\begin{equation}
 \Theta\mathcal H_{\mathrm{MF}}^{(\xi)}(h)\Theta^{-1}
 =\mathcal H_{\mathrm{MF}}^{(\xi)}(-h).
 \label{eq:appendix_theta_hamiltonian}
\end{equation}
It therefore becomes a symmetry inside either selected domain at \(h=0\).
Let
\(\bar{\bk}\equiv-R_6\bk\), folded back into the magnetic Brillouin zone, and
define \(\mathcal U_{\Theta}(\bk)\) by
\begin{equation}
 \Theta\Psi_{\bk,+}\Theta^{-1}
 =\mathcal U_{\Theta}(\bk)\Psi_{\bar{\bk},-}.
 \label{eq:appendix_theta_spinor}
\end{equation}
This matrix contains the site permutation, the magnetic-cell translation
phases, and the PSG gauge phases appearing in
Eq.~\eqref{eq:appendix_projective_c6}.  It maps particles to particles and
holes to holes, so
\(\mathcal U_{\Theta}^{\dagger}\tau_3\mathcal U_{\Theta}=\tau_3\).
Equation~\eqref{eq:appendix_theta_hamiltonian} is then equivalent to the
block identity
\begin{equation}
 \mathcal M_{\bar{\bk},-}^{(\xi)}(-h)
 =\mathcal U_{\Theta}^{\dagger}(\bk)
  \bigl[\mathcal M_{\bk,+}^{(\xi)}(h)\bigr]^*
  \mathcal U_{\Theta}(\bk).
 \label{eq:appendix_theta_matrix_relation}
\end{equation}
At \(h=0\), this relation pairs every positive-energy mode of the \(+\) block
with a mode of the \(-\) block having the same energy.  Because \(\Theta\) is
antiunitary and \(R_6\) is a proper rotation, their Berry
curvatures satisfy
\begin{equation}
 \varepsilon_{n\bar{\bk},-}^{(\xi)}
 =\varepsilon_{n\bk,+}^{(\xi)},
 \qquad
 \Omega_{n\bar{\bk},-}^{xy,(\xi)}
 =-\Omega_{n\bk,+}^{xy,(\xi)}.
 \label{eq:appendix_energy_berry_relation}
\end{equation}
Because the two partners have identical Bose weights, their contributions to
Eq.~\eqref{eq:kappa_kubo} cancel.  More generally,
\begin{equation}
 \kappa_{xy,-}^{(\xi)}(-h)
 =-\kappa_{xy,+}^{(\xi)}(h),
 \qquad
 \kappa_{xy}^{(\xi)}(-h)=-\kappa_{xy}^{(\xi)}(h),
 \label{eq:appendix_kappa_odd}
\end{equation}
where
\(\kappa_{xy}^{(\xi)}=\kappa_{xy,+}^{(\xi)}+
\kappa_{xy,-}^{(\xi)}\).  Setting \(h=0\) yields
\begin{equation}
 \kappa_{xy,+}^{(\xi)}(0)
 =-\kappa_{xy,-}^{(\xi)}(0),
 \qquad
 \kappa_{xy}^{(\xi)}(0)=0.
 \label{eq:appendix_zero_field_cancellation}
\end{equation}
Thus choosing one chiral domain does not select only one BdG block.  Both
\(\Psi_+\) and \(\Psi_-\) remain in that domain, and their zero-field Hall
responses cancel.  The numerical curves in
Fig.~\ref{fig:chiral_h0_cancellation}(b) evaluate the two terms independently.
Their sum vanishes to numerical precision over the full temperature grid.
A finite Zeeman field removes the internal
\(\Theta\) symmetry but retains the field-reversal relation in
Eq.~\eqref{eq:appendix_kappa_odd}, which accounts for the odd-in-field curves
in Fig.~\ref{fig:small_field}.

\section{Boson-Number Check at \texorpdfstring{$\kappa=1$}{kappa=1}}
\label{app:kappa_one}

\begin{figure}[t]
\centering
\includegraphics[width=\columnwidth]{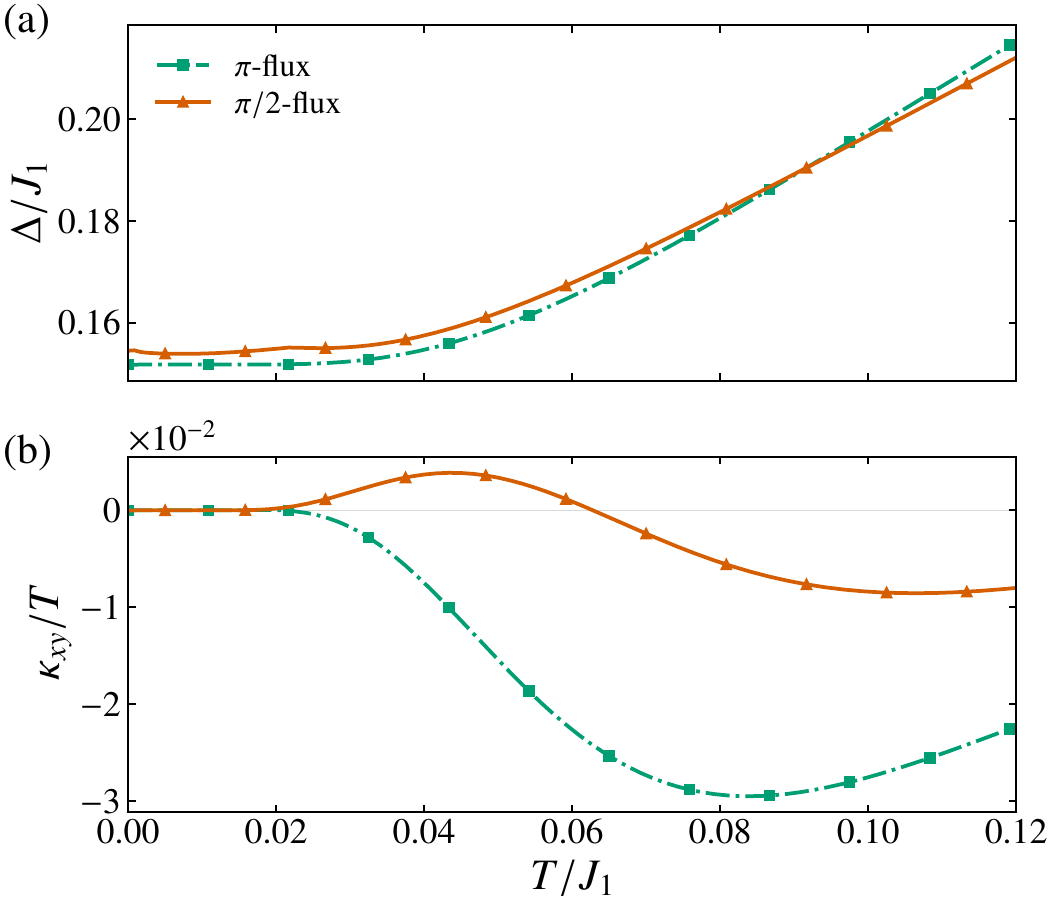}
\caption{
Results at \(\kappa=1\) for the uncondensed flux sectors at
\(J_2/J_1=0.20\), \(D/J_1=0.10\), and \(h/J_1=0.05\).  (a) Spinon gaps at
\(\kappa=1\).  (b) Corresponding intrinsic thermal Hall responses
\(\kappa_{xy}/T\).
}
\label{fig:kappa_one_appendix}
\end{figure}

We repeated the common-parameter calculation using the microscopic
spin-\(1/2\) boson-number value \(\kappa=2S=1\), while continuing to impose the
constraint only on average.  The \(\pi\)- and \(\pi/2\)-flux branches remain
gapped, with zero-temperature gaps of \(0.152J_1\) and \(0.155J_1\),
respectively.  By contrast, the \(0\)-flux gap closes before
\(\kappa=1\), signaling spinon condensation and magnetic order within
SBMFT.  This branch is therefore omitted from
Fig.~\ref{fig:kappa_one_appendix}, which compares only gapped, uncondensed
solutions.

\newpage
\begin{figure}[t]
\centering
\includegraphics[width=\columnwidth]{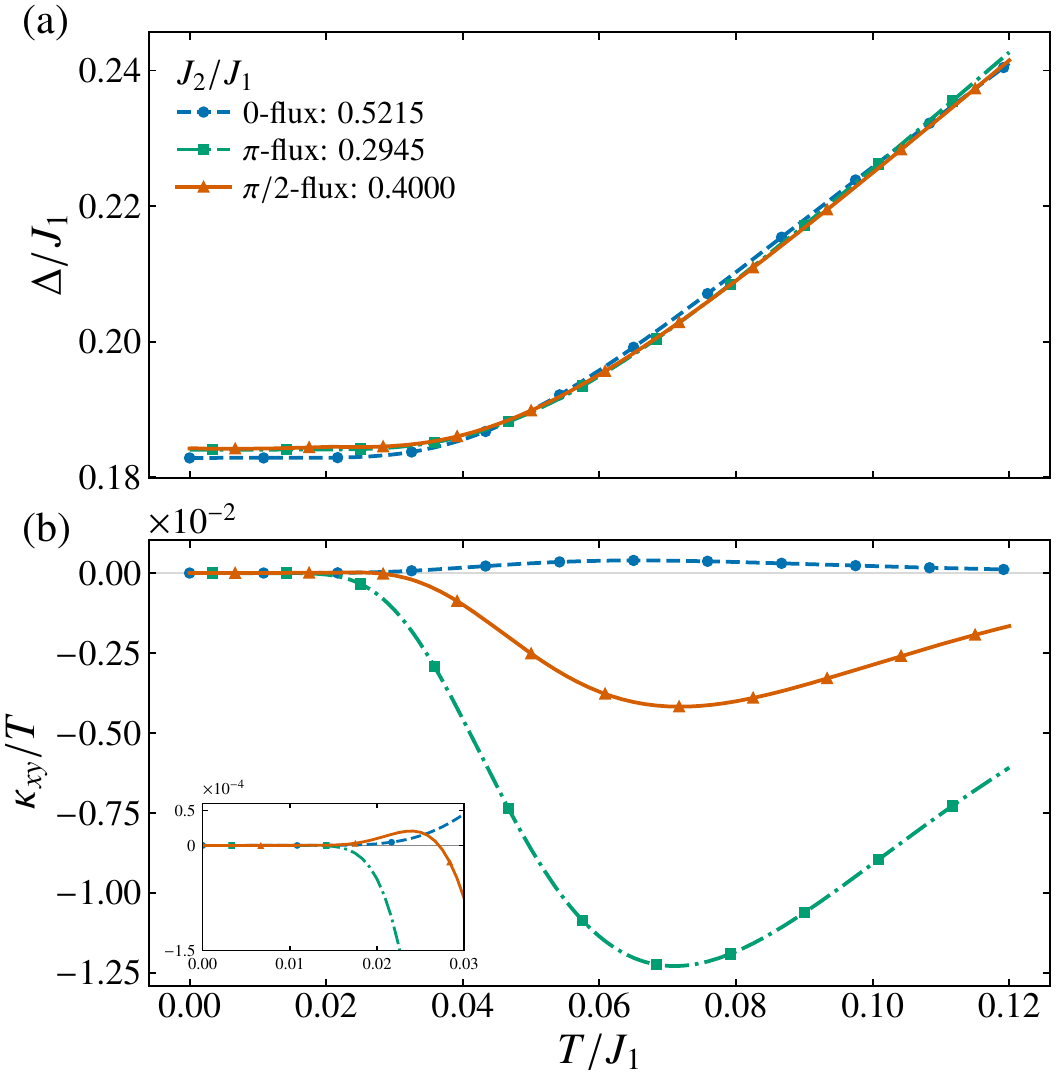}
\caption{
Auxiliary gap-matched control at \(D/J_1=0.03\) and \(h/J_1=0.05\).
(a) Spinon gaps for the \(0\)-, \(\pi\)-, and \(\pi/2\)-flux saddle-point
solutions at
\(J_2/J_1=\JtwoZero\), \(\JtwoPi\), and \(\JtwoPiHalf\), respectively.
(b) Corresponding intrinsic thermal Hall responses \(\kappa_{xy}/T\).  The inset enlarges the
low-temperature response near zero for all three sectors.
}
\label{fig:similar_gap_appendix}
\end{figure}

The relative fragility of the \(0\)-flux state is qualitatively consistent
with the earlier PSG-based mean-field analysis in Ref.~\cite{Wang-2010}.  In
that analysis, a spin-\(1/2\) zero-flux state is gapped only within a narrow
range of the ratio between next-nearest- and nearest-neighbor pairing fields,
whereas the nearest-neighbor \(\pi\)-flux state has a critical boson density
well above unity.  Figure~\ref{fig:kappa_one_appendix} shows the two
uncondensed branches retained here.  Both remain gapped throughout
\(0\leq T/J_1\leq0.12\).  The \(\pi\)-flux thermal Hall response becomes
negative as temperature increases, while the \(\pi/2\)-flux response is positive at
intermediate temperature and changes sign near \(T/J_1=0.063\).  At the
sum-rule-corrected value used in the main text, the corresponding crossover
occurs near \(T/J_1=0.0475\).  Thus the crossover position and response
magnitude vary with the constraint prescription, whereas the negative
\(\pi\)-flux response and sign-changing \(\pi/2\)-flux response persist for
the two gapped branches.

\section{Gap-Matched Control}
\label{app:similar_gap}

This comparison tests whether the flux dependence of the thermal Hall
response can be reduced to a difference in the lowest spinon gap.  The
Hamiltonian parameters are not identical across sectors.  The ratio \(J_2/J_1\) is
adjusted separately in each sector to match the three gap curves, while
\(D/J_1\) and \(h/J_1\) are held
fixed.  The gaps in Fig.~\ref{fig:similar_gap_appendix}(a) were matched at five
calibration temperatures and remain within \(0.9\%\) of one another over the
full plotted interval.  Their relative spread at \(T/J_1=0.08\) is
\(\GapSpread\), while the corresponding thermal Hall curves remain separated.
At the same temperature, the magnitude ratios of the thermal Hall
conductivities are
\(\lvert\kappa_{\pi}/\kappa_0\rvert=\KappaRatioPiZero\),
\(\lvert\kappa_{\pi/2}/\kappa_0\rvert=\KappaRatioPiHalfZero\), and
\(\lvert\kappa_{\pi/2}/\kappa_{\pi}\rvert=\KappaRatioPiHalfPi\).  The
\(\pi/2\)-flux curve of \(\kappa_{xy}/T\) has a positive maximum of only
\(2.05\times10^{-5}\) near \(T/J_1=0.024\) and changes sign near
\(T/J_1=0.027\).  This positive part is not resolved on the main-panel scale
set by the negative minimum of \(\kappa_{xy}/T\),
\(-4.18\times10^{-3}\), and is displayed in the inset.  The control therefore does
not establish that the response difference is caused by flux alone, because
the values of \(J_2\) differ, but it does show that matching the lowest spinon
gap does not collapse the three thermal Hall responses onto a common curve.

\bibliography{refs,refs_revision}

\end{document}


\title{Supplemental Material for ``Thermal Hall Signatures of Distinct
Schwinger-Boson Flux Sectors on the Honeycomb Lattice''}

\author{Daiki Sasamoto}
\email[sasamoto.daiki.r6@dc.tohoku.ac.jp]{}
\affiliation{
  Department of Physics, Graduate School of Science, Tohoku University,
  Sendai, Miyagi 980-8578, Japan
}

\date{October 25, 2026}

\maketitle

\section{Coordinate conventions}

Figure~1 of the main text shows the lattice and Brillouin-zone geometry.  For
reproducibility, we list here only the exact coordinates used to generate the
band and dynamical-structure-factor paths.  With nearest-neighbor distance
\(a=1\), our real- and reciprocal-space primitive vectors are
\begin{align}
\bm a_1&=\left(\sqrt{3},0\right),&
\bm a_2&=\left(\frac{\sqrt{3}}{2},\frac{3}{2}\right),\notag\\
\bm b_1&=\left(\frac{2\pi}{\sqrt{3}},-\frac{2\pi}{3}\right),&
\bm b_2&=\left(0,\frac{4\pi}{3}\right).
\label{eq:supp_primitive_vectors}
\end{align}
The high-symmetry points in the three reciprocal cells are
\begin{align}
0\text{-flux}:\quad
&\Gamma=\bm0,\quad
\mathrm M=\frac{\bm b_1+\bm b_2}{2},\quad
\mathrm K=\frac{2\bm b_1+\bm b_2}{3},
\notag\\
\pi\text{-flux}:\quad
&\bm g_1=\frac{\bm b_1}{2},\quad \bm g_2=\bm b_2,\quad
\mathrm X_{\pi}=\frac{\bm g_1}{2},\quad
\mathrm M_{\pi}=\bm g_1+\frac{\bm g_2}{2},
\notag\\
\pi/2\text{-flux}:\quad
&\bm g_1=\frac{\bm b_1}{4},\quad \bm g_2=\bm b_2,\quad
\mathrm X_{\pi/2}=\frac{\bm g_1}{2},\quad
\mathrm M_{\pi/2}=\frac{3\bm g_1+\bm g_2}{2}.
\label{eq:supp_high_symmetry_points}
\end{align}
For the extended physical-zone path used in the dynamical structure factor,
the equivalent zone center is \(\Gamma'=\bm b_1+\bm b_2\).

\section{Complete specification of the mean-field Ansätze}

\begin{figure}[t]
\centering
\includegraphics[width=\textwidth]{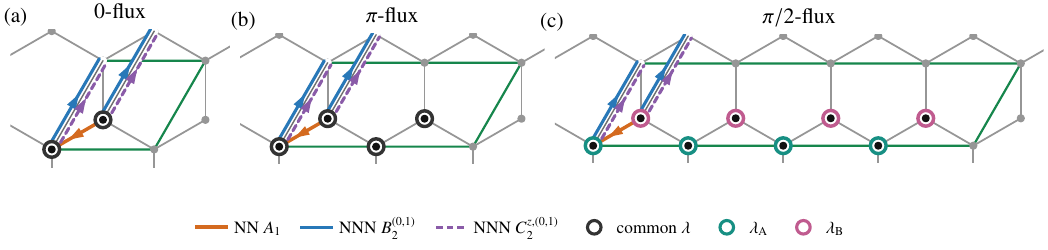}
\caption{
Independent saddle-point variables and magnetic unit cells used in the
mean-field calculations for the
(a) \(0\)-, (b) \(\pi\)-, and (c) \(\pi/2\)-flux Ansätze.  The green outline
marks the magnetic unit cell used in each mean-field calculation.  The orange
directed bond is the independent
NN pairing representative \(A_1\).  The parallel solid blue and dashed
purple arrows follow the same representative NNN links because the
\(B_2^{(0,1)}\) and
\(C_2^{z,(0,1)}\) components are two mean-field channels on those links, not
different physical bonds.  Black site rings denote the common multiplier
\(\lambda\) in the \(0\)- and \(\pi\)-flux Ansätze.  Teal and magenta rings
denote \(\lambda_{\mathrm A}\) and \(\lambda_{\mathrm B}\) in the
\(\pi/2\)-flux Ansatz.  Gray bonds provide lattice context.  All uncolored
site and bond variables are related to the displayed representatives by the
projective symmetries and are therefore not additional independent
coordinates.
}
\label{fig:supp_ansatz_parameterization}
\end{figure}

This section specifies the independent saddle-point coordinates and
the bonds generated from them by symmetry.  Magnetic translations and the
retained projective point-group operations fix the phase or complex
conjugation of each related bond.  Symmetry-related sites share one Lagrange
multiplier and one number constraint.  Table~\ref{tab:independent_coordinates}
lists the independent coordinates, and
Fig.~\ref{fig:supp_ansatz_parameterization} shows the corresponding bond
representatives, magnetic unit cells, and multipliers.

Let the magnetic unit cell used in the mean-field calculation contain sites
\(A_m\) and \(B_m\), with \(m=0,\ldots,q-1\), and take
\(\bm L_1=q\bm a_1\) and \(\bm L_2=\bm a_2\).  The values
\(q=1,2,4\) describe the 0-, \(\pi\)-, and \(\pi/2\)-flux sectors,
respectively.  Table~\ref{tab:independent_coordinates} lists all independent
real saddle-point coordinates used at finite field.  A complex representative
contributes two real coordinates.  Each listed saddle-point coordinate is
varied independently, although a mean-field amplitude may converge to zero.

\begin{table}[H]
\caption{
Independent saddle-point coordinates in the numerical Ansätze.  The four
complex NNN representatives are defined explicitly below.  The last column
counts independent real coordinates, including the Lagrange multipliers.
}
\label{tab:independent_coordinates}
\begin{ruledtabular}
\begin{tabular}{lccccc}
Flux sector & Magnetic unit cell
& Independent NN field & Independent NNN fields
& Multipliers & Real dimension \\
\hline
0-flux
& \(2\) sites, \(q=1\)
& \(A_1\in\mathbb R_{>0}\)
& \(B_2^{(0)},B_2^{(1)},C_2^{z,(0)},C_2^{z,(1)}\in\mathbb C\)
& \(\lambda\) & 10 \\
\(\pi\)-flux
& \(4\) sites, \(q=2\)
& \(A_1\in\mathbb R_{>0}\)
& \(B_2^{(0)},B_2^{(1)},C_2^{z,(0)},C_2^{z,(1)}\in\mathbb C\)
& \(\lambda\) & 10 \\
\(\pi/2\)-flux
& \(8\) sites, \(q=4\)
& \(A_1\in\mathbb R_{>0}\)
& \(B_2^{(0)},B_2^{(1)},C_2^{z,(0)},C_2^{z,(1)}\in\mathbb C\)
& \(\lambda_{\mathrm A},\lambda_{\mathrm B}\) & 11
\end{tabular}
\end{ruledtabular}
\end{table}

\subsection{Nearest-neighbor pairing pattern}

For each \(m\), the three directed nearest-neighbor bonds start at \(A_m\)
and end at \(B_m\), at the copy of \(B_m\) displaced by \(-\bm a_2\), and
at \(B_{m-1}\) along \(-\bm a_1\), respectively.  The index \(m-1\) is
understood modulo \(q\).  Their singlet fields are
\begin{equation}
 \mean{\mathcal A_m(0)}=A_1,
 \qquad
 \mean{\mathcal A_m(-\bm a_2)}=A_1e^{im\phi},
 \qquad
 \mean{\mathcal A_m(-\bm a_1)}=A_1,
 \label{eq:supp_nn_orbit}
\end{equation}
with \(\phi=0,\pi,\pi/2\) for \(q=1,2,4\), respectively.  The overall
U($1$) gauge is fixed by choosing \(A_1>0\).  Reversing a directed bond uses
\(\mathcal A_{ji}=-\mathcal A_{ij}\).

\subsection{Next-nearest-neighbor hopping patterns}

We retain positively directed NNN bonds in the directions
\(\bm a_1\), \(\bm a_2\), and \(\bm a_1-\bm a_2\) on both sublattices.
The DM signs in this orientation are
\begin{equation}
 \bigl(\nu_{\bm a_1},\nu_{\bm a_2},
 \nu_{\bm a_1-\bm a_2}\bigr)_{\mathrm A}=(-1,+1,+1),
 \qquad
 \bigl(\nu_{\bm a_1},\nu_{\bm a_2},
 \nu_{\bm a_1-\bm a_2}\bigr)_{\mathrm B}=(+1,-1,-1).
 \label{eq:supp_dm_signs}
\end{equation}

For a complex representative \(z\), define the primary symmetry-generated NNN pattern
\(\mathscr R_q(z,s,m,\bm\delta)\), where \(s=\mathrm A,\mathrm B\), by
\begin{align}
 \mathscr R_1(z,s,m,\bm\delta)={}&z,
 \label{eq:supp_R1}\\
 \mathscr R_2(z,s,m,\bm\delta)={}&
 \begin{cases}
 z, & \bm\delta=\bm a_1,\\
 (-1)^m z, & \bm\delta=\bm a_2,\\
 (-1)^m z, & \bm\delta=\bm a_1-\bm a_2, s=\mathrm A,\\
 -(-1)^m z, & \bm\delta=\bm a_1-\bm a_2, s=\mathrm B,
 \end{cases}
 \label{eq:supp_R2}\\
 \mathscr R_4(z,s,m,\bm\delta)={}&
 \begin{cases}
 z, & \bm\delta=\bm a_1,\\
 i^m z^*, & \bm\delta=\bm a_2, s=\mathrm A,\\
 (-i)^m z^*, & \bm\delta=\bm a_1-\bm a_2, s=\mathrm A,\\
 (-i)^m z^*, & \bm\delta=\bm a_2, s=\mathrm B,\\
 i^{m+1}z^*, & \bm\delta=\bm a_1-\bm a_2, s=\mathrm B.
 \end{cases}
 \label{eq:supp_R4}
\end{align}
For the positive \(\pi/2\)-flux domain, we define
\(\mathscr R_4^{(+)}\equiv\mathscr R_4\), with \(\mathscr R_4\) given by
Eq.~\eqref{eq:supp_R4}.  The conjugate domain is obtained from
\begin{equation}
 \mathscr R_4^{(-)}(z,s,m,\bm\delta)
 =\left[\mathscr R_4^{(+)}(z^*,s,m,\bm\delta)\right]^*.
 \label{eq:supp_conjugate_orbit}
\end{equation}

Define the sign factor
\begin{equation}
 \chi_q(s,m,\bm\delta)=
 \begin{cases}
 \nu_{s,m,\bm\delta}, & q=1,2,\\
 -1, & q=4,\ s=\mathrm A,\\
 +1, & q=4,\ s=\mathrm B.
 \end{cases}
 \label{eq:supp_cross_character}
\end{equation}
The complete NNN fields entering the BdG Hamiltonian are then
\begin{align}
 \mean{\mathcal B_{s,m}(\bm\delta)}={}&
 \mathscr R_q\bigl(B_2^{(0)},s,m,\bm\delta\bigr)
 +\chi_q(s,m,\bm\delta)
 \mathscr R_q\bigl(B_2^{(1)},s,m,\bm\delta\bigr),
 \label{eq:supp_B_orbit}\\
 \mean{\mathcal C^z_{s,m}(\bm\delta)}={}&
 \chi_q(s,m,\bm\delta)
 \mathscr R_q\bigl(C_2^{z,(0)},s,m,\bm\delta\bigr)
 +\mathscr R_q\bigl(C_2^{z,(1)},s,m,\bm\delta\bigr).
 \label{eq:supp_C_orbit}
\end{align}
The reverse NNN orientation follows by Hermitian conjugation.  The second
representative in each channel is the additional finite-field symmetry component.
Setting \(B_2^{(1)}=C_2^{z,(1)}=0\) gives the restricted Ansatz rather than the
Ansatz used for the reported data.

\subsection{Channels not used as saddle coordinates}

The formal bond-operator basis contains
\(\mathcal A,\mathcal B,\mathcal C^z,\mathcal D^z\), but the numerical
branches compared in the main text use the independent coordinates listed in
Table~\ref{tab:independent_coordinates}.  In particular,
\(B_1,C_1^z,D_1^z,A_2\), and \(D_2^z\) are not varied as independent
saddle coordinates.  The NNN DM interaction is nevertheless retained.  Within
this branch it enters the quadratic Hamiltonian through the
\(B_2\)-\(C_2^z\) cross decoupling.  The omitted channels are evaluated after
convergence as diagnostics.  The converged solutions give
\(B_1,C_1^z,A_2,D_2^z=0\) within numerical precision.  A finite \(D_1^z\)
correlator may be induced in a fixed chiral \(\pi/2\)-flux domain, but it is
not fed back as a variational coordinate because there is no NN DM coupling.
Thus the comparison is performed in the same explicitly specified
\(A_1\)-\(B_2\)-\(C_2^z\) bond-channel manifold for all three flux sectors.

Finally, the 0- and \(\pi\)-flux sectors use one common multiplier \(\lambda\).
The fixed \(\pi/2\)-flux domain has two inequivalent sets of symmetry-related
sites and uses \(\lambda_{\mathrm A}\) and \(\lambda_{\mathrm B}\), with the
number constraint imposed separately on the two sets.  Its zero-temperature
parameters are defined by the continuous \(T\to0^+\) limit of these
site-resolved saddle-point equations.

\Needspace{0.78\textheight}
\section{\texorpdfstring{\red{Berry curvature and the \(\pi/2\)-flux sign reversal}}{Berry curvature and the pi/2-flux sign reversal}}
\label{sec:supp_berry_maps}

\begin{figure}[H]
\centering
\red{\includegraphics[width=0.66\textwidth]{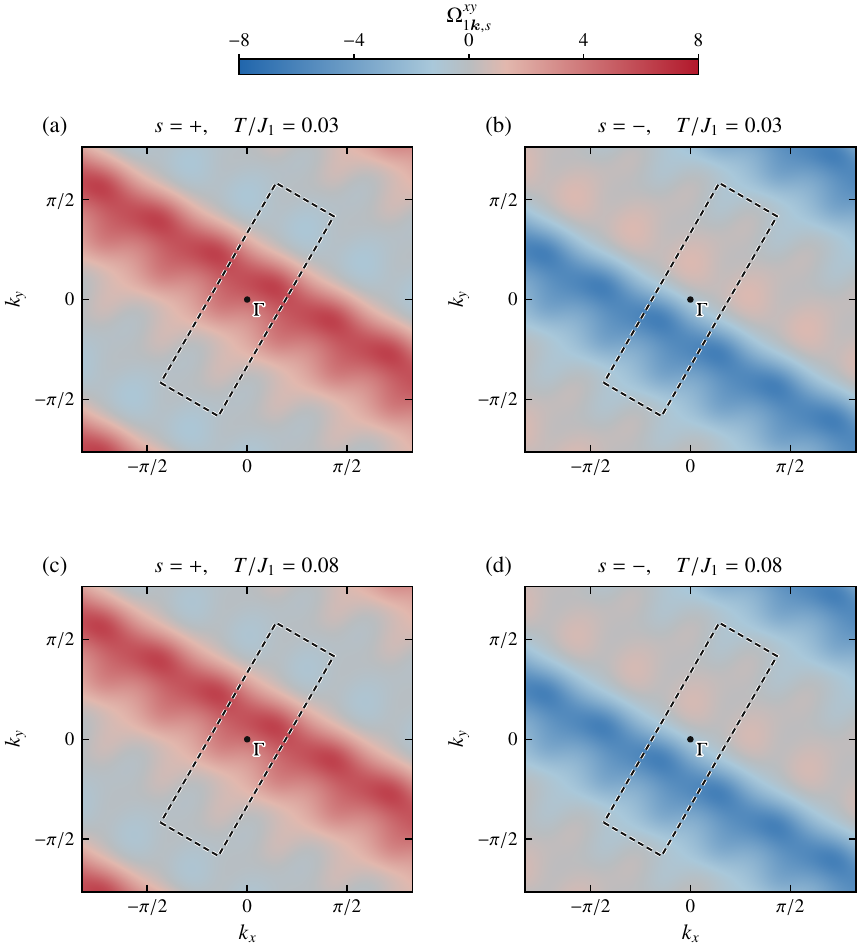}}
\caption{\red{Berry-curvature maps for the \(\pi/2\)-flux saddle at
\(\kappa=\sqrt{3}-1\), \(J_2/J_1=0.20\), \(D/J_1=0.10\), and
\(h/J_1=0.05\).
The total \(\kappa_{xy}/T\) is positive at \(T/J_1=0.03\) in (a),(b)
and negative at \(T/J_1=0.08\) in (c),(d).
The left and right columns show the lowest positive-energy band of the
U($1$) blocks \(s=+\) and \(s=-\), respectively.
The dashed boundary encloses the magnetic reciprocal primitive cell of
Fig.~1(d) of the main text, and the surrounding area shows its periodic
continuation.}}
\label{fig:supp_pi2_lowest_band_maps}
\end{figure}

\red{Figure~\ref{fig:supp_pi2_lowest_band_maps} complements the
high-symmetry-path results in Fig.~5 of the main text with momentum-space maps
before and after the \(\pi/2\)-flux sign reversal.  We show the lowest
positive-energy band in each U($1$) block.  These maps are unweighted,
whereas the thermal Hall response includes all positive-energy bands
with their thermal weights.}

To resolve the temperature-driven sign reversal, we define the
thermal-kernel-weighted Berry-curvature density
\begin{equation}
 I_T(\varepsilon)=-\frac{1}{A_{\mathrm M}N_k^2}
 \sum_{n,\bm k}
 \left[c_2\!\left(n_{\mathrm B}(\varepsilon_{n\bm k})\right)
 -\frac{\pi^2}{3}\right]
 \Omega^{xy}_{n\bm k}\,
 \delta(\varepsilon-\varepsilon_{n\bm k}),
 \label{eq:supp_thermal_integrand}
\end{equation}
so that
\begin{equation}
 \frac{\kappa_{xy}}{T}=\int d\varepsilon\,I_T(\varepsilon).
 \label{eq:supp_thermal_integral}
\end{equation}
\Needspace{4\baselineskip}
Here \(A_{\mathrm M}\) is the magnetic-unit-cell area.  The numerical
evaluation uses the same positive-norm bands, analytic momentum derivatives,
and Berry curvature as the transport calculation in the main text.

\begin{figure}[H]
\centering
\includegraphics[width=\textwidth]{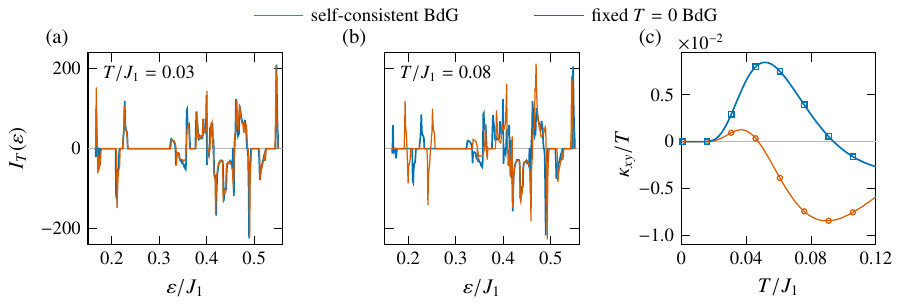}
\caption{
Energy-resolved origin of the \(\pi/2\)-flux sign reversal at
\(J_2/J_1=0.20\), \(D/J_1=0.10\), and \(h/J_1=0.05\).
(a),(b) Weighted density \(I_T(\varepsilon)\) at \(T/J_1=0.03\) and
\(0.08\), respectively.  Solid orange curves use the self-consistent BdG
Hamiltonian at the displayed temperature.  Solid blue curves hold the
\(T=0\) BdG Hamiltonian fixed and vary only the Bose kernel.
(c) Corresponding \(\kappa_{xy}/T\) over the full temperature range.
The energy integrals in (a) and (b) reproduce the corresponding values of
\(\kappa_{xy}/T\) obtained independently from the band-resolved Kubo formula.
}
\label{fig:supp_pi2_sign_reversal}
\end{figure}

For the self-consistent Hamiltonian, the energy integral is positive at
\(T/J_1=0.03\) and negative at \(T/J_1=0.08\), on opposite sides of the
crossing in Fig.~\ref{fig:supp_pi2_sign_reversal}(c).  The fixed-\(T=0\)
control remains positive across this crossing region.  Changing the Bose
occupation factor alone therefore does not reproduce the observed crossing.
The sign reversal at the self-consistent crossing temperature therefore
requires the temperature-dependent reconstruction of the band energies,
Bogoliubov eigenvectors, and Berry-curvature distribution.
\label{para:supp_crossing_stability_start}
\red{The tested \(\pi/2\)-flux crossing temperatures remain unchanged at
the precision quoted in the main text under variations of the numerical
settings.}
\label{para:supp_crossing_stability_end}

\section{\texorpdfstring{\red{Noncondensed ranges and \(J_2\) dependence}}{Noncondensed ranges and J2 dependence}}
\label{sec:supp_j2_window}

\red{At \(\kappa=\sqrt{3}-1\), \(D/J_1=0.10\), and \(h/J_1=0.05\),
we extend the mean-field sweep to \(0\leq J_2/J_1\leq1\) at
\(T/J_1=0,0.02,0.04,0.08,0.12\).
Table~\ref{tab:supp_uncondensed_ranges} summarizes the gapped self-consistent
solutions obtained at the sampled points.
At \(T=0\), gapped \(\pi\)- and \(\pi/2\)-flux solutions are obtained
down to \(J_2=0\).  For the \(0\)-flux branch, gapped solutions are obtained
at the sampled points with \(0.1703\leq J_2/J_1\leq1\), and the lower
endpoint has a spinon gap of \(8.52\times10^{-4}J_1\).}

\begin{table}[H]
\caption{\red{Verified \(J_2/J_1\) ranges of gapped, uncondensed saddles,
with rounded endpoints.
The last column is the common range at all five temperatures listed in the text.}}
\label{tab:supp_uncondensed_ranges}
\begin{ruledtabular}
\red{\begin{tabular}{ccc}
Flux sector & \(T=0\) & All five temperatures \\
\hline
\(0\) & \([0.1703,1.00]\) & \([0.175,1.00]\) \\
\(\pi\) & \([0,1.00]\) & \([0,1.00]\) \\
\(\pi/2\) & \([0,1.00]\) & \([0,1.00]\)
\end{tabular}}
\end{ruledtabular}
\end{table}

\begin{figure}[H]
\centering
\red{\includegraphics[width=0.48\textwidth]{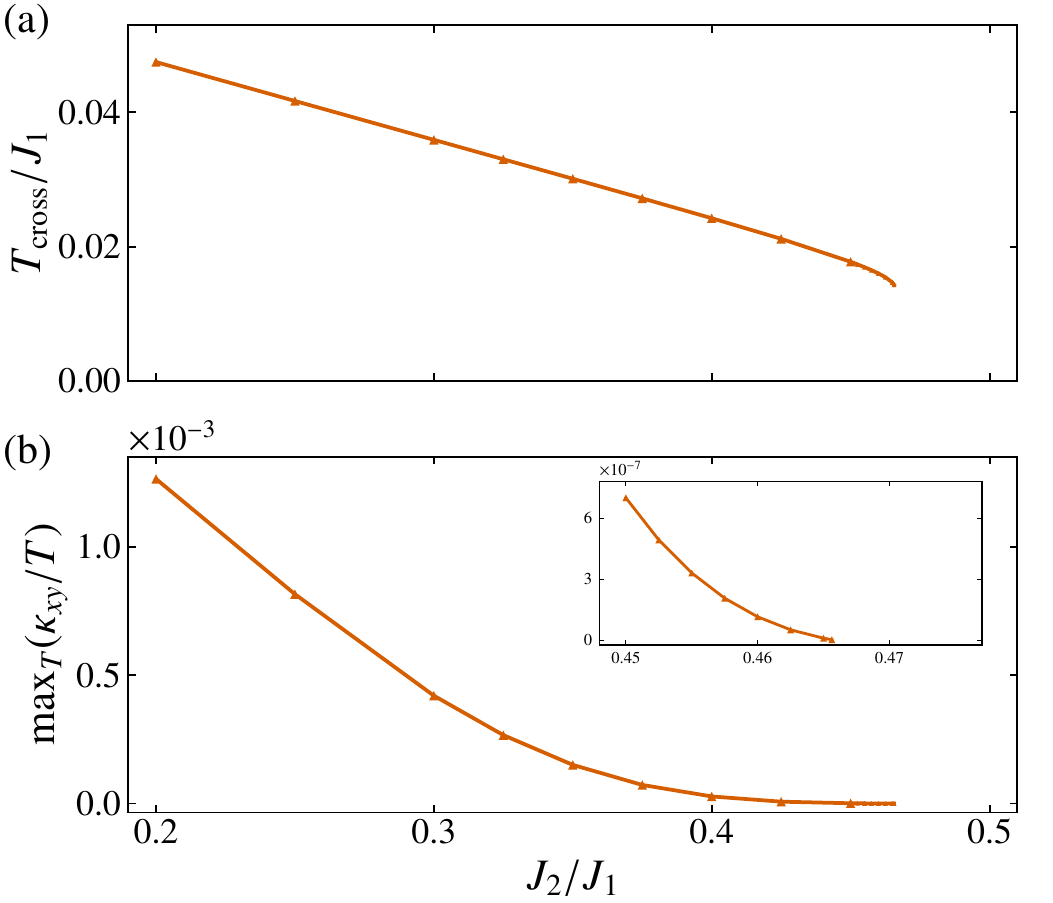}}
\caption{\red{\(J_2\) dependence of the \(\pi/2\)-flux response at
\(\kappa=\sqrt{3}-1\), \(D/J_1=0.10\), and \(h/J_1=0.05\).
(a) Upper positive-to-negative crossing temperature and (b) maximum of the
positive low-temperature \(\kappa_{xy}/T\) window.
The inset enlarges the small positive peaks.}}
\label{fig:supp_pi2_j2_window}
\end{figure}

\red{Both the crossing temperature and the positive peak decrease with
\(J_2/J_1\) (Fig.~\ref{fig:supp_pi2_j2_window}).  The upper endpoint marks
the last numerically resolved positive window and does not establish a sharp
boundary for the sign reversal.}

\Needspace{8\baselineskip}
\section{\texorpdfstring{\red{Ordered-state magnon thermal Hall benchmark}}{Ordered-state magnon thermal Hall benchmark}}
\label{sec:supp_magnon_benchmark}

\red{We examine linear spin waves of the same spin Hamiltonian and DM-arrow
convention as in the main text, with physical spin \(S=1/2\).
At the common spinon-comparison point \(J_2/J_1=0.20\),
\(D/J_1=0.10\), \(h/J_1=0.05\), the N\'eel reference is dynamically
unstable, with an imaginary frequency reaching \(0.251J_1\).
A lower-energy conical reference is harmonically stable, but its
zero-point spin reduction is \(\delta S\simeq1.65>S\), invalidating its bare
harmonic ordered-moment
expansion.  We therefore use low-\(J_2\) N\'eel states as illustrative
spin-wave benchmarks, not as predictions for the ordered state at the common
parameter point.
These checks neither exclude other magnetic order nor establish a
spin-liquid ground state.}

\red{For the field-canted two-sublattice N\'eel reference, the unit spin
directions on A and B are \((\sin\theta,0,\cos\theta)\) and
\((-\sin\theta,0,\cos\theta)\), respectively, with canting angle fixed by}
\begin{equation}
 \red{\cos\theta=\frac{h}{6J_1S}.}
 \label{eq:supp_neel_reference}
\end{equation}
\red{Expanding in local Holstein--Primakoff bosons and retaining quadratic
terms gives two positive-energy magnon bands.  Magnon thermal Hall transport
in the \(J_2=0\) limit is discussed in Ref.~\cite{Owerre-2017-Neel-Hall}.
For the bond convention used here, the lattice structure factors are}
\begin{align}
 \red{\gamma_{\bm k}}&\red{=1+e^{-i\bm k\cdot\bm a_1}
                       +e^{-i\bm k\cdot\bm a_2},}\notag\\
 \red{F_{\bm k}}&\red{=\cos(\bm k\cdot\bm a_1)
                  +\cos(\bm k\cdot\bm a_2)
                  +\cos[\bm k\cdot(\bm a_1-\bm a_2)],}\notag\\
 \red{G_{\bm k}}&\red{=-\sin(\bm k\cdot\bm a_1)
                  +\sin(\bm k\cdot\bm a_2)
                  +\sin[\bm k\cdot(\bm a_1-\bm a_2)].}
 \label{eq:supp_neel_structure_factors}
\end{align}
\red{The two magnon dispersions are}
\begin{equation}
 \red{\varepsilon_{\pm,\bm k}^2
 =S^2\Bigl\{\Bigl[3J_1+2J_2(F_{\bm k}-3)
 \pm\sqrt{J_1^2\cos^4\theta|\gamma_{\bm k}|^2
              +4D^2\cos^2\theta G_{\bm k}^2}\Bigr]^2
 -J_1^2\sin^4\theta|\gamma_{\bm k}|^2\Bigr\}.}
 \label{eq:supp_neel_dispersion}
\end{equation}
\red{The Goldstone mode \(\varepsilon_{-,\Gamma}=0\) is retained without
adding a gap.  Berry curvatures are evaluated from the paraunitary modes,
including both positive- and negative-energy intermediate states and the
physical orbital positions in the momentum derivatives.  The A
site is at the origin and the B site at \((\bm a_1+\bm a_2)/3\).
With the same thermal-current convention as the main text,}
\begin{equation}
 \red{\frac{\kappa_{xy}}{T}=-\frac{k_{\mathrm B}^2}{\hbar}
 \sum_{n=1}^{2}\int_{\mathrm{BZ}}\frac{d^2k}{(2\pi)^2}
 \left[c_2\!\left(n_{\mathrm B}(\varepsilon_{n\bm k})\right)
       -\frac{\pi^2}{3}\right]\Omega^{xy}_{n\bm k}.}
 \label{eq:supp_magnon_hall}
\end{equation}
\red{This per-layer intrinsic harmonic response is evaluated on a fixed
classical background.  It does not establish finite-temperature long-range
order in the strictly two-dimensional model.}

\red{For \(D/J_1=0.10\), the references at \(J_2/J_1=0,0.05,0.10\)
are harmonically stable at both fields in Table~\ref{tab:supp_magnon_benchmark}.
The zero-point spin reduction \(\delta S\) is the mean local
Holstein--Primakoff boson occupation per spin.
Although \(S-\delta S>0\) at these points, the reduction is already large
at \(J_2/J_1=0.10\).  A positive harmonic ordered moment is only a
necessary check, not a guarantee of quantitative control.
At \(h/J_1=0.05\), \(\delta S\) exceeds \(S\) at the sampled
\(J_2/J_1=0.125\) point (\(\delta S\simeq0.569\)).
At \(h/J_1=0.05\), the computed
response is odd under reversal of \(h\) or \(D\), and the \(D=0\) and
\(h=0\) controls vanish within numerical precision.}

\begin{table}[H]
\caption{\red{Low-\(J_2\) magnon benchmarks at \(S=1/2\) and
\(D/J_1=0.10\).  The last three columns give \(\kappa_{xy}/T\) in units
of \(k_{\mathrm B}^2/\hbar\) at the stated \(T/J_1\).}}
\label{tab:supp_magnon_benchmark}
\begin{ruledtabular}
\red{\begin{tabular}{cccccc}
\(J_2/J_1\) & \(h/J_1\) & \(\delta S\)
& \(T/J_1=0.03\) & \(0.08\) & \(0.12\) \\
\hline
0.00 & 0.05 & 0.251 & \(3.06\times10^{-7}\) & \(6.86\times10^{-6}\) & \(2.06\times10^{-5}\) \\
0.05 & 0.05 & 0.319 & \(8.93\times10^{-7}\) & \(1.82\times10^{-5}\) & \(4.57\times10^{-5}\) \\
0.10 & 0.05 & 0.446 & \(4.65\times10^{-6}\) & \(6.84\times10^{-5}\) & \(1.20\times10^{-4}\) \\
0.00 & 0.20 & 0.233 & \(4.01\times10^{-7}\) & \(2.48\times10^{-5}\) & \(1.21\times10^{-4}\) \\
0.05 & 0.20 & 0.293 & \(1.18\times10^{-6}\) & \(7.27\times10^{-5}\) & \(3.21\times10^{-4}\) \\
0.10 & 0.20 & 0.404 & \(6.23\times10^{-6}\) & \(3.19\times10^{-4}\) & \(1.04\times10^{-3}\)
\end{tabular}}
\end{ruledtabular}
\end{table}

\begin{figure}[H]
\centering
\red{\includegraphics[width=0.98\textwidth]{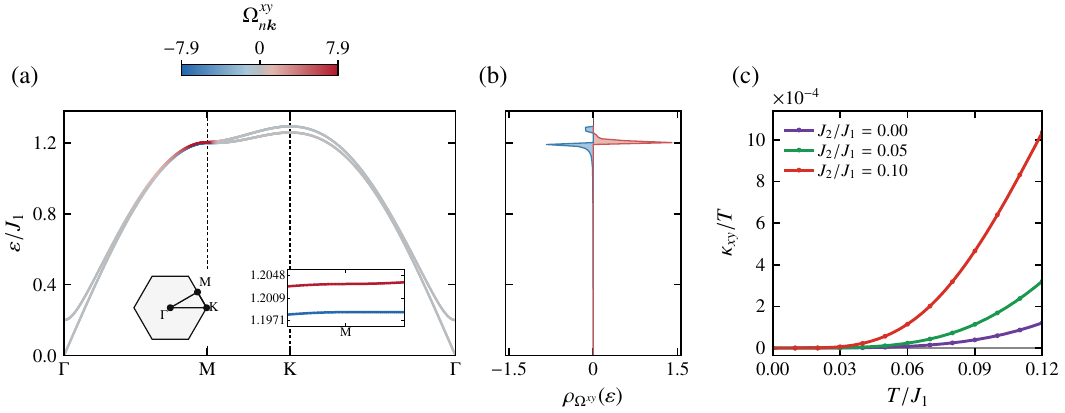}}
\caption{\red{Magnon thermal Hall benchmark at \(S=1/2\),
\(D/J_1=0.10\), and \(h/J_1=0.20\).
(a) Positive-energy N\'eel magnon bands at \(J_2/J_1=0.05\), colored by
the unweighted Berry curvature.  The insets show the physical
Brillouin-zone path and the band splitting near M.
(b) Corresponding unweighted signed Berry-curvature density.
(c) \(\kappa_{xy}/T\) at \(J_2/J_1=0,0.05,0.10\), in units of
\(k_{\mathrm B}^2/\hbar\).  The larger field resolves the small band
separation.  The response at the main-text field \(h/J_1=0.05\) is given in
Table~\ref{tab:supp_magnon_benchmark}.}}
\label{fig:supp_neel_magnon_benchmark}
\end{figure}

\red{The resolved magnon responses at these low-\(J_2\) points are positive
and increase over the displayed temperature interval, without a sign reversal.
This behavior is specific to the tested references and is not a universal
statement about magnon transport.}

\Needspace{20\baselineskip}
\section{Dynamical structure factor comparison}

We calculate the dynamical structure factor as a complementary probe of
the three mean-field states.  Momentum- and energy-resolved spin spectra are
widely used to characterize candidate spin-liquid states
\cite{Knolle-2014,Messio-Cepas-2010,Ralko-Merino-2024}.  This observable is
evaluated from the same converged saddles but does not enter the thermal
Hall calculation.

For \(\alpha=x,y,z\), we define the connected dynamical structure factor as
\begin{equation}
 S^{\alpha\alpha}(\bm q,\omega)
 =\frac{1}{N}\sum_{i,j}\int_{-\infty}^{\infty}\frac{dt}{2\pi}\,
 e^{i\omega t}e^{-i\bm q\cdot(\bm r_i-\bm r_j)}
 \left[
 \mean{S_i^\alpha(t)S_j^\alpha(0)}
 -\mean{S_i^\alpha}\mean{S_j^\alpha}
 \right],
 \label{eq:supp_dssf_definition}
\end{equation}
where \(N\) is the total number of lattice sites and \(\bm r_i\) is the
position of site \(i\).  The expectation value is taken in the bosonic BdG
vacuum \(\lvert0_{\mathrm{MF}}\rangle\) of the converged zero-temperature
saddle-point solution.  The spin operator evolves according to the fixed
quadratic mean-field Hamiltonian,
\begin{equation}
 S_i^\alpha(t)
 =e^{i\mathcal H_{\mathrm{MF}}t}S_i^\alpha
 e^{-i\mathcal H_{\mathrm{MF}}t},
 \qquad
 \mean{\mathcal O}
 =\langle0_{\mathrm{MF}}\rvert\mathcal O\lvert0_{\mathrm{MF}}\rangle.
 \label{eq:supp_dssf_time_evolution}
\end{equation}
The spin-summed response is
\begin{equation}
 S(\bm q,\omega)
 =\sum_{\alpha=x,y,z}S^{\alpha\alpha}(\bm q,\omega).
 \label{eq:supp_dssf_spin_sum}
\end{equation}
Its frequency and momentum integral is the equal-time onsite connected
correlator,
\begin{equation}
 \frac{1}{N}\sum_{\bm q}\int_{-\infty}^{\infty}d\omega\,
 S(\bm q,\omega)
 =\frac{1}{N}\sum_i\sum_{\alpha=x,y,z}
 \left[
 \mean{S_i^\alpha S_i^\alpha}
 -\mean{S_i^\alpha}\mean{S_i^\alpha}
 \right].
 \label{eq:supp_dssf_sum_rule}
\end{equation}
For the sum-rule-restoring choice \(\kappa=\sqrt{3}-1\) used in these
calculations, adding the disconnected elastic contribution at \(\omega=0\)
gives the spin-\(1/2\) moment sum rule \(3/4\).  The quadratic mean-field ground state is Gaussian,
so Wick's theorem gives an exact factorization within this state.  Applied to
the connected correlator in Eq.~\eqref{eq:supp_dssf_definition}, it gives
\begin{align}
 \mean{S_i^\alpha(t)S_j^\alpha(0)}
 -\mean{S_i^\alpha}\mean{S_j^\alpha}
 ={}&\frac14
 \sigma^\alpha_{\mu\nu}\sigma^\alpha_{\rho\lambda}
 \Bigl[
 \mean{b_{i\mu}^\dagger(t)b_{j\rho}^\dagger(0)}
 \mean{b_{i\nu}(t)b_{j\lambda}(0)}
 \notag\\[-2pt]
 &\hspace{20mm}+
 \mean{b_{i\mu}^\dagger(t)b_{j\lambda}(0)}
 \mean{b_{i\nu}(t)b_{j\rho}^\dagger(0)}
 \Bigr],
 \label{eq:supp_dssf_wick}
\end{align}
where repeated spin indices are summed.  Both anomalous and normal
contractions, including all sites in the magnetic unit cell, are retained.
The calculation uses the converged zero-temperature paraunitary eigenvectors
and a normalized Lorentzian broadening width
\(\delta_\omega/J_1=0.01\).

\begin{figure}[H]
\centering
\includegraphics[width=\textwidth]{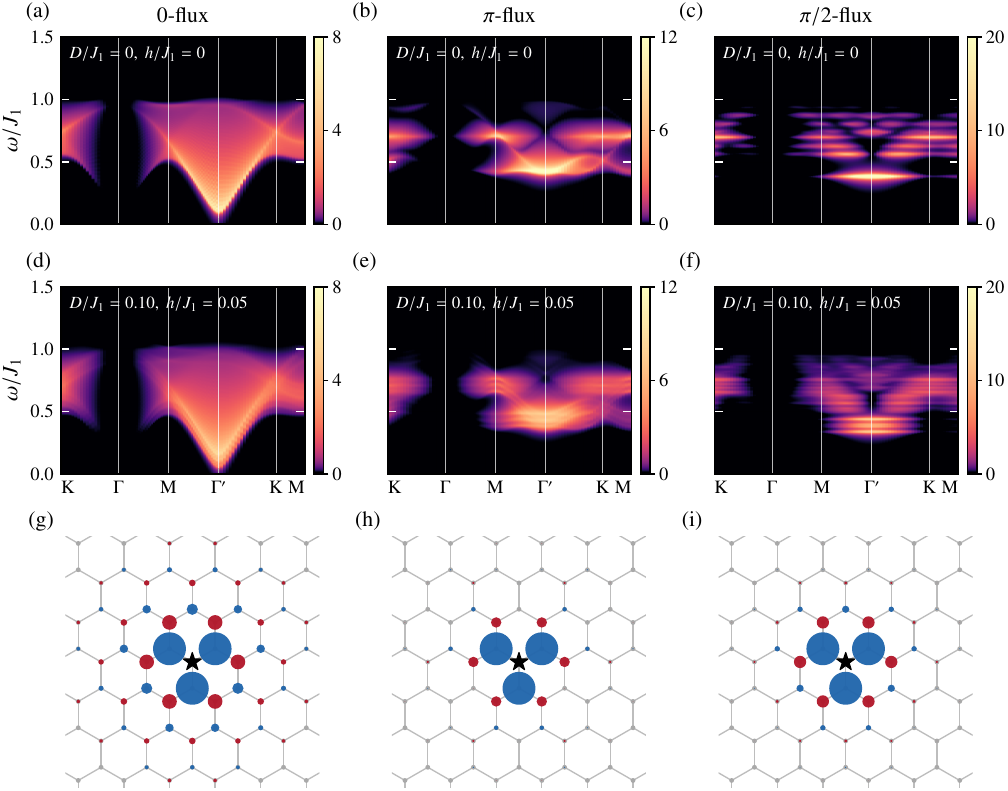}
\caption{
Connected zero-temperature dynamical structure factor
\(S(\bm q,\omega)=\sum_{\alpha=x,y,z}S^{\alpha\alpha}(\bm q,\omega)\)
along the \(\mathrm K\)--\(\Gamma\)--\(\mathrm M\)--\(\Gamma'\)--\(\mathrm K\)--\(\mathrm M\)
path.  The columns show the \(0\)-, \(\pi\)-, and \(\pi/2\)-flux Ansätze.
(a)--(c) use \(D/J_1=h/J_1=0\), while (d)--(f) use
\(D/J_1=0.10\) and \(h/J_1=0.05\).  Here \(J_2/J_1=0.20\) and
\(\delta_\omega/J_1=0.01\).  The intensity scale is shared vertically within
each column.  Panels (g)--(i) show connected equal-time real-space spin
correlations at the parameters of (d)--(f), measured from the reference site
(star) through the tenth-neighbor shell.  Blue and red circles denote negative
and positive correlations, respectively, and their areas are proportional to
the correlation magnitude normalized within each panel.
}
\label{fig:supp_dynamical_structure_factor}
\end{figure}

For the three zero-field calculations and for the finite-field \(0\)- and
\(\pi\)-flux calculations in Fig.~\ref{fig:supp_dynamical_structure_factor},
the local moments vanish within numerical precision, so the connected
response exhausts the full moment sum rule.  The finite-field \(\pi/2\)-flux
solution instead has a small sublattice-staggered local moment with zero net
magnetization.  The connected spectrum then excludes the associated
zero-frequency disconnected elastic contribution.  In every case, an
independent Brillouin-zone momentum and frequency sum of the connected spectrum
agrees with the equal-time connected correlator in
Eq.~\eqref{eq:supp_dssf_sum_rule}.  With this choice of \(\kappa\), restoring
the disconnected contribution numerically recovers the moment sum rule
\(3/4\) for all six calculations.

\Needspace{7\baselineskip}
The \(0\)-flux spectrum has a strongly dispersive lower boundary, whereas the
\(\pi\)-flux spectrum has a comparatively flat finite-energy onset and the
\(\pi/2\)-flux spectrum contains several nearly dispersionless ridges.  The
\(0\)-flux equal-time correlations remain appreciable over longer distances,
while the finite-flux correlations decrease more rapidly beyond the first few
shells.  This qualitative correspondence is consistent with the connection
between spectral flatness and short-ranged correlations found for a chiral
bosonic Ansatz of the spin-\(1\) Kitaev model
\cite{Sasamoto-Ralko-Merino-Nasu-2026}.

\Needspace{12\baselineskip}
\bibliography{refs,refs_revision}